%% file: main.tex
\documentclass[aip,reprint]{revtex4-1}
\usepackage[utf8]{inputenc}
\usepackage{fullpage}
\usepackage{amssymb}
\usepackage{graphicx}
\usepackage{lipsum}
\usepackage{hyperref}
\usepackage[dvipsnames]{xcolor}
\definecolor{myblue}{RGB}{25, 161, 224}
\definecolor{myblue}{RGB}{28, 54, 184}
\hypersetup{colorlinks, citecolor=blue, linkcolor=myblue, urlcolor=blue}
\usepackage{amsmath}
\usepackage{braket}
\usepackage{tablefootnote}
\usepackage{dcolumn}

\draft 

\begin{document}


\title{Directional detection of dark matter using solid-state quantum sensing} 




\author{Reza Ebadi}
\email[]{ebadi@umd.edu}
\affiliation{Department of Physics, University of Maryland, College Park, Maryland 20742, USA}
\affiliation{Quantum Technology Center, University of Maryland, College Park, Maryland 20742, USA}
\author{Mason C. Marshall}
\email[]{mason.marshall@nist.gov}
\affiliation{Quantum Technology Center, University of Maryland, College Park, Maryland 20742, USA}
\affiliation{Department of Electrical and Computer Engineering, University of Maryland, College Park, Maryland 20742, USA}
\author{David F. Phillips}
\affiliation{Center for Astrophysics $\vert$ Harvard \& Smithsonian, Cambridge, Massachusetts, 02138, USA}
\author{Johannes Cremer}
\affiliation{Quantum Technology Center, University of Maryland, College Park, Maryland 20742, USA}
\affiliation{Department of Electrical and Computer Engineering, University of Maryland, College Park, Maryland 20742, USA}
\affiliation{Department of Physics, Harvard University, Cambridge, Massachusetts 02138, USA}
\author{Tao Zhou}
\affiliation{Center for Nanoscale Materials, Argonne National Laboratory, Lemont, Illinois 60439, USA}
\author{Michael Titze}
\affiliation{Sandia National Laboratories, Albuquerque, NM, 87123, USA}
\author{Pauli Kehayias}
\affiliation{Sandia National Laboratories, Albuquerque, NM, 87123, USA}
\author{Maziar Saleh Ziabari}
\affiliation{Sandia National Laboratories, Albuquerque, NM, 87123, USA}
\author{Nazar Delegan}
\affiliation{Center for Molecular Engineering, Argonne National Laboratory, Lemont, Illinois 60439, USA}
\affiliation{Materials Science Division, Argonne National Laboratory, Lemont, Illinois 60439, USA}
\author{Surjeet Rajendran}
\affiliation{Department of Physics and Astronomy, Johns Hopkins University, 3400 N. Charles St., Baltimore, Maryland 21218, USA}
\author{Alexander O. Sushkov}
\affiliation{Department of Physics, Boston University, Boston, Massachusetts 02215, USA}
\affiliation{Department of Electrical and Computer Engineering, Boston University, Boston, Massachusetts 02215, USA}
\affiliation{Photonics Center, Boston University, Boston, Massachusetts 02215, USA}
\author{F. Joseph Heremans}
\affiliation{Center for Molecular Engineering, Argonne National Laboratory, Lemont, Illinois 60439, USA}
\affiliation{Materials Science Division, Argonne National Laboratory, Lemont, Illinois 60439, USA}
\affiliation{Pritzker School of Molecular Engineering, University of Chicago, Chicago, Illinois 60637, USA}
\author{Edward S. Bielejec}
\affiliation{Sandia National Laboratories, Albuquerque, NM, 87123, USA}
\author{Martin V. Holt}
\affiliation{Center for Nanoscale Materials, Argonne National Laboratory, Lemont, Illinois 60439, USA}
\author{Ronald L. Walsworth}
\email[]{walsworth@umd.edu}
\affiliation{Department of Physics, University of Maryland, College Park, Maryland 20742, USA}
\affiliation{Quantum Technology Center, University of Maryland, College Park, Maryland 20742, USA}
\affiliation{Department of Electrical and Computer Engineering, University of Maryland, College Park, Maryland 20742, USA}


\date{\today}

\begin{abstract}
Next-generation dark matter (DM) detectors searching for weakly interacting massive particles (WIMPs) will be sensitive to coherent scattering from solar neutrinos, demanding an efficient background-signal discrimination tool. Directional detectors improve sensitivity to WIMP DM despite the irreducible neutrino background. Wide-bandgap semiconductors offer a path to directional detection in a high-density target material. A detector of this type operates in a hybrid mode. The WIMP or neutrino-induced nuclear recoil is detected using real-time charge, phonon, or photon collection. The directional signal, however, is imprinted as a durable sub-micron damage track in the lattice structure. This directional signal can be read out by a variety of atomic physics techniques, from point defect quantum sensing to x-ray microscopy. In this Review, we present the detector principle as well as the status of the experimental techniques required for directional readout of nuclear recoil tracks.  Specifically, we focus on diamond as a target material; it is both a leading platform for emerging quantum technologies and a promising component of next-generation semiconductor electronics. Based on the development and demonstration of directional readout in diamond over the next decade, a future WIMP detector will leverage or motivate advances in multiple disciplines towards precision dark matter and neutrino physics.
\end{abstract}


\maketitle 

\clearpage
\newpage
\tableofcontents
\input{intro}
\input{sections}
\input{conclusion}

\acknowledgments We wish to acknowledge substantial contributions and collaborations of Matthew J Turner and Connor Hart in various aspects of the project. We would like to acknowledge technical discussions and assistance from Mark J H Ku and Raisa Trubko. This work was supported by the Argonne National Laboratory under Award No. 2F60042;the DOE QuANTISED program under Award No. DE-SC0019396; the Army Research Laboratory MAQP program under Contract No. W911NF-19-2-0181; the DARPA DRINQS program under Grant No. D18AC00033; the DOE fusion program under Award No. DE-SC0021654; and the University of Maryland Quantum Technology Center. Work performed at the Center for Nanoscale Materials and Advanced Photon Source, both U.S. Department of Energy Office of Science User Facilities, was supported by the U.S. DOE, Office of Basic Energy Sciences, under Contract No. DE-AC02-06CH11357. This work was performed, in part, at the Center for Integrated Nanotechnologies, an Office of Science User Facility operated for the U.S. Department of Energy (DOE) Office of Science. Sandia National Laboratories is a multimission laboratory managed and operated by National Technology \& Engineering Solutions of Sandia, LLC, a wholly owned subsidiary of Honeywell International, Inc., for the U.S. DOE’s National Nuclear Security Administration under contract DE-NA-0003525. The views expressed in the article do not necessarily represent the views of the U.S. DOE or the United States Government.\\
{\bf Data Availability.} The data that support the findings of this study are available from the corresponding author upon reasonable request.\\
{\bf Conflict of interest.} The authors have no conflicts to disclose.
\bibliography{references}

\end{document}

%% file: intro.tex
\section{Introduction}
\label{sec_intro}
The nature of dark matter (DM) is one of the most pressing puzzles in modern physics \cite{PDG}. Possible candidates for DM include weakly interacting massive particles (WIMPs) in the mass range of a few GeV to 100 TeV. Thermally-produced WIMP DM acquires its observed relic abundance through the freeze-out mechanism in the early universe \cite{WIMP_production_2015_BAER}. WIMPs are also strongly motivated from the perspective of model building; they naturally arise in numerous theories beyond the Standard Model \cite{WIMP_theory_2006_LSP,WIMP_theory_2007_extraDim_Hooper,WIMP_theory_2014_littleHiggs,WIMP_theory_2015_peskin_SUSY,WIMP_theory_2021_nearNuFloor_Bottaro}. For decades, direct detection experiments have sought to observe WIMP-induced nuclear recoils in a background-controlled environment; their null results to date have excluded a large portion of possible WIMP cross section-mass parameter space \cite{WIMP_DD_2019_Schumann}.

Next-generation multiton experiments will be sensitive to even lower cross sections \cite{darwin_2016,argo_2018}. However, as their sensitivity increases, they will face an irreducible background: coherent neutrino scattering \cite{snowmass_CEvNS}. This background, traditionally called the ``neutrino floor'', produces nuclear recoil spectra similar to those of WIMP-induced recoils \cite{spectra_2007,spectra_2009,spectra_2010,spectra_2014}. Thus, additional information is required to distinguish between the neutrino background signal and the putative WIMP signal. The disparate angular distributions of WIMP and neutrino fluxes can provide such a discrimination tool. Due to the motion of the solar system within the Galactic halo, a WIMP DM flux would exhibit a dipolar structure \cite{Spergel_1988}. Solar neutrinos, the most significant neutrino background \cite{solar_nu_search_xe_2021}, also feature a relatively small angular spread. Directional detectors, which would determine the incoming direction of incident particles, can therefore distinguish between WIMP- and neutrino-induced nuclear recoils if they have sufficient sensitivity and angular resolution \cite{directional_2014,directional_2015,Mayet:2016zxu,Vahsen:2020pzb,Vahsen:2021gnb}. As a result, multiple efforts to develop directional detectors are underway worldwide, each with its own advantages and challenges \cite{Vahsen:2021gnb}. 

We envision a solid-state detector for directional WIMP detection that operates in a hybrid mode: nuclear recoil event registration is performed in real time using charge, phonon, or photon collection; and directional information is read out using high-resolution mapping of the stable damage left by the incident WIMPs or neutrinos in the detector's crystal lattice \cite{Rajendran:2017ynw,Marshall:2020azl}. The proposed detector scheme integrates real-time event registration, three-dimensional directional information, and high-density target material. Gas-based time-projection chambers (TPCs) have mature directional detection technology and offer real-time event detection and full 3D directional readout as well \cite{Vahsen:2020pzb,snowmass_recoilImaging}; however, they have low target density and thus require very large volumes. The scalability of these detectors is therefore limited. Probing DM mass $\rm{>10\,GeV}$ below the atmospheric neutrino floor, despite being theoretically well-motivated \cite{WIMP_thoery_highM_atmNu2011_Hisano,WIMP_thoery_highM_atmNu2014_Roszkowski,WIMP_thoery_highM_atmNu2016_Abdullah,WIMP_thoery_highM_atmNu2017_GAMBIT,WIMP_thoery_highM_atmNu2018_Kobakhidze,Asadi:2021bxp,Asadi:2022vkc}, would be especially challenging for gas TPCs \cite{OHare_atmospheric_2020}. In contrast, solid-state detectors could allow to access this parameter space.

In the proposed detector, directional readout and event registration require different sets of technologies. DM detection using nuclear or electron recoils in semiconductors is a mature technology; semiconductor-based DM detection experiments are already operational \cite{SuperCDMS_LopezAsamar:2019smu,SuperCDMS_Rau:2020abt,DAMIC:2021crr,DAMIC:2021esz}. Our proposed experiment will use similar methods to register real-time recoil events \cite{Rajendran:2017ynw,Marshall:2020azl}. Mapping of directional lattice damage, however, requires a different set of capabilities, currently under development. Our research is focused on developing directional detectors enabled by point defect quantum sensing in wide-bandgap semiconductors, possibly augmented by x-ray microscopy. We discuss in the following sections how recent developments in defect-based quantum sensing pave the way for directional detection. We specifically focus on diamond as a viable target material. Diamond is a leading platform for emerging quantum technologies \cite{sensitivity_Barry2020}, and over the past decade many technical developments have been made in diamond-based quantum sensing that are applicable to directional readout. In addition, the demand for quantum-grade synthetic diamonds has resulted in efficient protocols for the production of high-quality, uniform-crystalline diamonds \cite{strain_control_2009_friel,strain_lowdislocations_typeIIa_2009_martineau_high}\footnote{Recent studies suggest that geologically old rock samples can also be used as particle detectors to detect damage tracks induced by ultraheavy dark matter (see, e.g., \cite{Hardy:2014mqa,Gresham:2017cvl,Grabowska:2018lnd,Carney:2022gse}) using electron microscopy \cite{Ebadi:2021cte}. In addition, it is envisioned to detect WIMP- or neutrino-induced damage tracks in natural minerals \cite{Snowden-Ifft1995,Collar:1994mj,engel1995,SnowdenIfft:1997hd,Baum:2018tfw,Drukier:2018pdy,Edwards:2018hcf,Baum:2019fqm,Baum:2021chx,Baum:2021jak,Baum:2022wfc,Jordan:2020gxx}. These methods provide large exposures with relatively small target masses since the signal accumulation time could be more than a billion years (see also \cite{Acevedo:2021tbl,Bramante:2021dyx}).}.

Implementing conventional DM detection methods in diamond is also a topic of ongoing research with promising prospects \cite{Kurinsky:2019pgb,Canonica:2020omq,snowmass_lowThreshold}. By combining directional readout methods utilizing quantum sensing \cite{Rajendran:2017ynw,Marshall:2020azl,sensitivity_Barry2020} and nuclear recoil registration methods (similar to those used in silicon and germanium-based detectors \cite{SuperCDMS_LopezAsamar:2019smu,SuperCDMS_Rau:2020abt,DAMIC:2021crr,DAMIC:2021esz}), it would be possible to build a detector that retains sensitivity to WIMPs despite the presence of an irreducible neutrino background.

The rest of the paper is organized as follows. In Sec.\,\ref{sec_nufloor}, we discuss the neutrino background for direct DM detectors; briefly discuss directional detection, current and proposed technologies; and give an overview of the working principle of the envisioned solid-state detector. In Sec.\,\ref{sec_nv}, we present an overview of nitrogen-vacancy centers in diamond and relevant quantum sensing techniques. In Sec.\,\ref{sec_methods}, we summarize the state-of-the-art techniques for directional readout in diamond, summarize our recent advances, and discuss anticipated improvements. A method of producing injected signals for characterization of detector efficiency is discussed in Sec.\,\ref{sec_ion_implantation}. We conclude with a summary and outlook in Sec.\,\ref{sec_discussion}. Readers interested only in a high-level overview of the project can refer to Sec.\,\ref{sec_discussion}.

%% file: sections.tex
\section{WIMP detection below the neutrino floor}
\label{sec_nufloor}
\subsection{Neutrino floor}
\label{sec_neutrinofloor_subsection}

Neutrinos couple to quarks via the neutral Z boson exchange \cite{GargamelleNeutrino_1973}; at neutrino energies below a few tens of MeV, this allows coherent scattering from atomic nuclei \cite{Freedman_coherent_1974,snowmass_CEvNS}. With technological advances in detecting low-energy (down to a few keV) nuclear recoils, coherent elastic neutrino-nucleus scattering (CE$\nu$NS) has been observed recently in CsI[Na] scintillator \cite{COHERENT_1_2017} and liquid argon \cite{COHERENT_2_2021} detectors by the COHERENT collaboration \cite{COHERENT_snowmass}. Next-generation multi-ton-scale DM direct detection experiments will also be sensitive to CE$\nu$NS from solar and atmospheric neutrinos, which cannot be shielded against, forming an irreducible background known as the {\it neutrino floor}. In fact, xenon-based DM detectors (e.g., XENONnT) are expected to detect CE$\nu$NS from ${\rm ^8B}$ solar neutrinos in the near future, which would be analogous to the detection of $6~{\rm GeV}\,c^{-2}$ DM \cite{solar_nu_search_xe_2021}. DARWIN is a next-generation liquid xenon time projection chamber (TPC) detector proposed by the XENON collaboration, which employs a 50-ton (30-ton) active (fiducial) mass and can operate for five years; it aims for sensitivity at the atmospheric neutrino floor \cite{darwin_2016}. In addition, the DarkSide collaboration has proposed ARGO, a 300-ton (200-ton) active (fiducial) liquid argon TPC detector with a five-year operation time, with a goal also to achieve sensitivity set by the neutrino floor \cite{argo_2018}.

There can be a substantial overlap between the recoil spectra induced by neutrinos and WIMPs, which makes distinguishing them difficult \cite{spectra_2007,spectra_2009,spectra_2010,spectra_2014}. A small difference in the tails of the recoil spectra could in principle be used as a discrimination tool, but it would require thousands of neutrino events, making it impractical with any proposed future experiment. Other methods for circumventing the neutrino floor have been proposed, such as utilizing timing information \cite{timing_2015,timing_2021} or complementarity between detectors \cite{complementarity_2014}. Nevertheless, in the low-statistics limit, directional detection appears to be the only feasible method \cite{directional_2014,directional_2015}.

\begin{figure}[htbp]
\begin{center}
\includegraphics[width=0.48\textwidth]{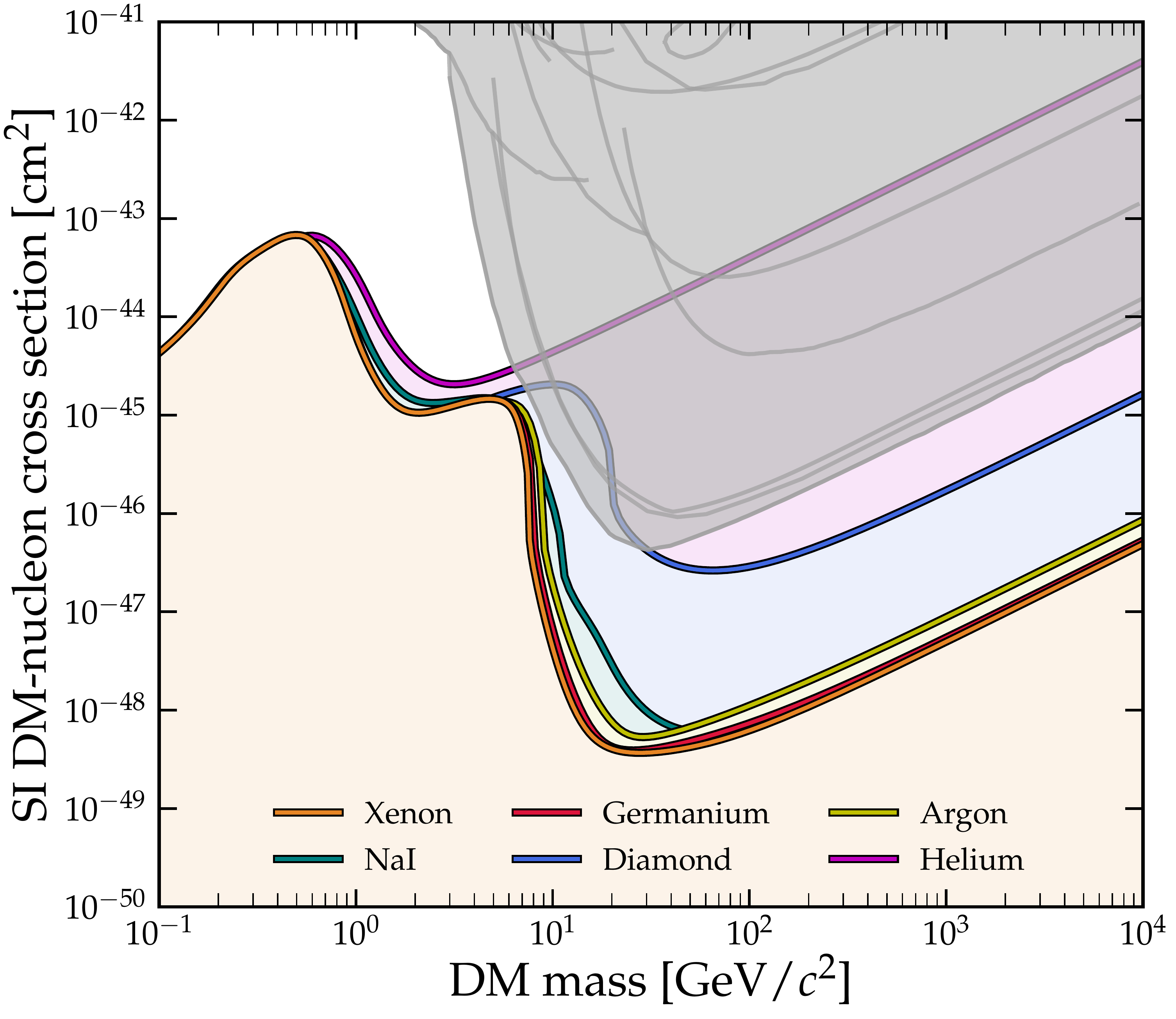}
\end{center}
\caption{\label{fig:neutrino_fog} Spin-independent DM-nucleon cross section versus DM mass parameter space, illustrating the neutrino floor for a variety of target materials, including diamond (using the typically dominant $^{12}$C isotope and produced using the method introduced in ref.\,\cite{nu_fog_2021}). Shaded gray area shows current excluded region and gray lines within this region represent limits from different experiments. Refer to ref.\cite{nu_fog_2021} for further discussion.}
\end{figure}

Despite the similarities between the signals, a conventional DM search would have been possible with a precise understanding of neutrino background levels. Recently some have advocated using the term ``neutrino fog'' rather than ``neutrino floor'', implying a challenging, but not impossible signal-background discrimination \cite{complementarity_2014,timing_2015,timing_2021,nu_fog_Gelmini:2018ogy,nu_fog_Gaspert:2021gyj,nu_fog_2021}. Even though neutrino-nucleus cross sections are relatively accurately determined, neutrino fluxes are typically subject to large systematic uncertainties, causing large uncertainties in the number of expected neutrino events \cite{theory_uncertainty_2021}. Thus, the neutrino flux uncertainty saturates the DM discovery limit for higher exposures \cite{spectra_2014}. Ref.\,\cite{nu_fog_2021} defines the neutrino floor in terms of the derivative of the discovery limit as a function of exposure $n=-({\rm d}\ln\sigma/{\rm d}\ln N_\nu)^{-1}$, where $N_\nu$ is the number of neutrino events and it is proportional to the exposure; the neutrino floor is the boundary leaving the standard Poissonian-statistics regime ($n=2$) and beginning systematic uncertainty saturation ($n>2$). (Note that while this definition of neutrino floor is useful for illustrating the neutrino background and comparing different target materials, it does not represent a fundamentally preferred definition.) Figure\,\ref{fig:neutrino_fog} shows the neutrino floor calculated using the method described above for several different target materials. Solar neutrinos dominate the neutrino floor for DM masses $\gtrsim~5~{\rm GeV}$; for slightly larger masses, the diffuse supernova neutrino background dominates; and eventually, for DM masses over a few tens of GeV, atmospheric neutrinos are the dominant contribution \cite{nu_fog_2021}. For light target nuclear masses like helium or carbon,  solar neutrinos are also a significant contributor in this higher DM mass regime since the scattering kinematics allow for large recoil energies \cite{nu_fog_2021}. As a result, a higher neutrino floor is expected at heavier DM mass ranges for light target nuclei (such as carbon in diamond). However, we do not expect this to significantly reduce the sensitivity of a diamond-based directional DM detector due to an efficient rejection of the solar neutrino background through leveraging localized angular distributions of the solar neutrino flux; see Sec.\,\ref{sec_directional_nuclear_recoil_detection}.

\subsection{Directional nuclear recoil detection}
\label{sec_directional_nuclear_recoil_detection}

Since the solar system moves towards the constellation Cygnus in the Galactic rest frame, the angular distribution of the DM scattering rate is expected to exhibit a dipolar feature in the lab frame \cite{Spergel_1988}. The angular distance between Cygnus and the sun is always at least 60$^{\circ}$ (varying between 60$^{\circ}$ and 90$^{\circ}$ over the course of a year). This angular separation allows the separation of solar neutrinos and dark matter in a directional nuclear recoil detector \cite{directional_2015,Mayet:2016zxu,Vahsen:2020pzb,Vahsen:2021gnb,snowmass_recoilImaging}. The dipolar DM angular structure can improve DM reach beyond atmospheric neutrino floor as well; however, to achieve sensitivity to atmospheric neutrinos, higher exposure is required \cite{Vahsen:2020pzb}. Although the input atmospheric neutrino flux is not localized as solar neutrino flux is, it exhibits angular structure caused by cosmic rays passing through different thicknesses of atmosphere and their interaction with geomagnetic fields \cite{OHare_atmospheric_2020}. Such a feature could further enhance the efficiency of directional detection.

In addition, local dark matter phase-space structures could be directly observed using directional detectors \cite{directdetection_MW_debrisflow_2012,OHare:2018trr}. There is overwhelming evidence that Milky Way kinematic structures go beyond the equilibrium Standard Halo Model \cite{Newberg2002,Crane2003,Gomez2012,Widrow2012,Gomez2013,Carlin2013,Williams2013,Xu2015,Carrillo2018,Schoenrich2018,Antoja2018,Laporte2019,Bennett2019,Necib2019,Necib2019a,MW_DM_2021}. Characterizing local DM structure is an active field of research, with implications on interpretation of direct detection experiments \cite{Vogelsberger:2008qb,Buch:2019aiw,Evans:2018bqy,Ibarra:2018yxq,Phillips:2020xmf}. Directional detectors that provide more kinematic information about incident particles can improve our understanding of local DM properties \cite{directdetection_MW_debrisflow_2012,OHare:2018trr}.  

Several directional detection technologies are in various stages of development~\cite{Vahsen:2021gnb}. Identifying three-dimensional direction and head/tail information on an event-by-event basis is the ideal scenario for directional detection. Gas-phase time projection chamber (TPC) detectors \cite{NEWAGE_2010,MIMAC_2011,D3_2012,DRIFT_2014,CYGNO_2020} provide well-developed 3D vectorial directional readout capabilities \cite{Vahsen:2020pzb}. However, gas-phase detectors that operate at and below the neutrino floor would require extremely large volumes, since the total target mass determines the detector's discovery limit. The use of nuclear emulsion-based detectors provides high-resolution recoil track imaging at higher target densities; however, they are time-consuming to read out and provide only time-integrated signals (hence, a detector installed on a Cygnus-tracking equatorial telescope would be optimal) \cite{NEWS_2016,NEWSdm_2017,emulsion_2018,NEWSdm_2020,emulsion_2020_directional60keV,emulsion_2021}. Although emulsion-based detectors can image 2D recoil tracks, it is unclear whether they can reconstruct 3D tracks and head/tail signatures. 

There are also indirect signatures of DM directionality that can be detected. Solid-state anisotropic scintillators (e.g., ${\rm ZnWO_4}$) \cite{scintillator_2002,scintillator_2003} can exploit modulation of incoming DM direction relative to crystal axes in order to statistically provide a directional signal \cite{scintillator_2005_diurnal,scintillator_2013,scintillator_2016_ADAMO,scintillator_2020_ADAMO} (although strong anisotropy at low recoil energies still has to be confirmed experimentally). A statistical measurement of this kind would require much higher exposures than event-resolved directional methods. As a directional signal, the relative direction of the primary ionization cloud and the applied electric field affects the scintillation yield in liquid noble gas detectors, a method known as columnar recombination \cite{columnar_recombination_2013,columnar_recombination_2014}. The first measurements of direction dependence of scintillation yield, however, suggests a small effect. Also, this method is sensitive to only one track dimension and lacks a head/tail signature, so it is probably ineffective for DM detection below the neutrino floor \cite{OHare_atmospheric_2020}. 

We envision using quantum point defects and possibly x-ray diffraction microscopy in solid-state detectors to provide directional readout. Due to intensive work on instrumentation of sensitive charge and phonon detectors during the last decade, semiconductor DM detectors based on silicon or germanium have become possible \cite{SuperCDMS:2005vdw,SuperCDMS_Speller:2015mna,SuperCDMS:2016wui,SuperCDMS_LopezAsamar:2019smu,SuperCDMS_Rau:2020abt,DAMIC:2021crr,DAMIC:2021esz}. According to the same principles and using developed instrumentation technologies, wide-bandgap semiconductors such as diamond \cite{Rajendran:2017ynw,Kurinsky:2019pgb} and silicon carbide \cite{Rajendran:2017ynw,Griffin:2020lgd} have been proposed for DM detection. These detectors' semiconductor properties and low nuclear mass provide a complementary sensitivity profile to existing detectors \cite{Kurinsky:2019pgb}. Furthermore, a WIMP or neutrino event in such a detector would leave a stable track of crystal lattice damage of length $\rm{\sim100\,nm}$, with the crystal thereby acting as a ``frozen bubble chamber'' to record the direction of the particle's impact \cite{Rajendran:2017ynw,Marshall:2020azl}. The WIMP or neutrino-induced damage track would result from a cascade of secondary nuclear recoils triggered by the initial scattering. The orientation and head/tail direction of this damage track could be mapped in three dimensions via spectroscopy of quantum point defects such as nitrogen-vacancy (NV) \cite{NV_review_2014_Schirhagl,NV_review_Levine_2019,NV_review_2020_edmonds,sensitivity_Barry2020} and silicon-vacancy (SiV) \cite{SiV_2018} defects in diamond, and divacancies in silicon carbide \cite{divacancy_SiC_2018_Awschalom,divacancy_SiC_2018_Gali}. X-ray diffraction microscopy could also be used as a complementary or alternative tool to map the WIMP or neutrino-induced damage track. (In Section~\ref{sec_methods}, we cover readout methods in greater detail.) Simulations for a diamond target indicate measurable orientation and head/tail asymmetry down to 1-3 keV of recoil energy \cite{Rajendran:2017ynw}. Therefore, quantum defect-based solid-state detectors offer full directional information (akin to gas TPCs), as well as high density.

\subsection{Working principle of a solid-state quantum defect directional detector}
\label{sec_detector_principle}

\begin{figure*}[tbp]
\begin{center}
\includegraphics[width=0.85\textwidth]{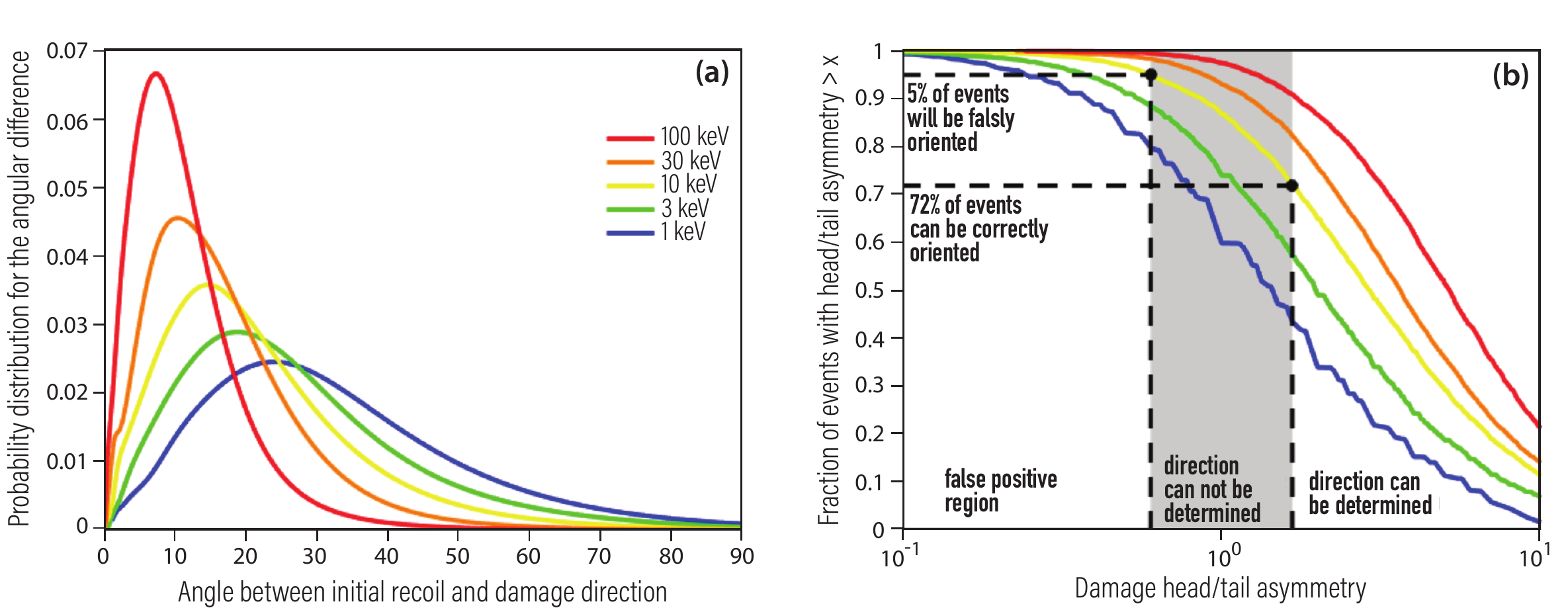}
\end{center}
\caption{\label{fig:SRIM} Result of a SRIM simulation assuming a carbon lattice (appropriate for diamond) and implantation of a carbon ion with energy in the range $\sim 1-100\,{\rm keV}$ (representing the initial recoiling nucleus induced by an incoming WIMP or neutrino). {\bf (a)} Distribution of the angular difference between the initial recoil direction and the damage track direction. Higher recoil energies are predicted to have a higher correlation. {\bf (b)} Distribution of damage head/tail asymmetry. The asymmetry is defined as the ratio of the number of lattice vacancies and interstitial nuclei in the first and last third of the damage. As an example, for ${\rm 10\,keV}$ initial ion energy, we predict about 70\% efficiency and 5\% false positive rate. Reprinted with permission from S. Rajendran, N. Zobrist, A. O. Sushkov, R. Walsworth, and M. Lukin, Phys. Rev. D 96, 035009 (2017). Copyright (2017) by the American Physical Society.}
\end{figure*}

In a solid-state crystalline detector, a WIMP recoil would impart substantial kinetic energy onto a target nucleus, knocking it off from its lattice site. This recoiling nucleus would initiate a chain of secondary recoils (via conventional standard-model interactions), leaving behind a characteristic damage track of interstitial nuclei, lattice vacancies, and distorted bonds. SRIM simulations \cite{SRIM} for a diamond detector predict damage tracks will be tens of nanometers long for recoils with energies in the range of ${\rm 10 - 100\,keV}$ \cite{Rajendran:2017ynw}, equivalent to WIMP masses in the range of ${\rm 1 - 100\,GeV}$; the orientation of these damage tracks will be well correlated with the incoming WIMP direction and initial recoil orientation (see Figure\,\ref{fig:SRIM}a); and the tracks will exhibit an observable head/tail asymmetry (see Figure\,\ref{fig:SRIM}b).

\begin{figure}[htbp]
\begin{center}
\includegraphics[width=0.48\textwidth]{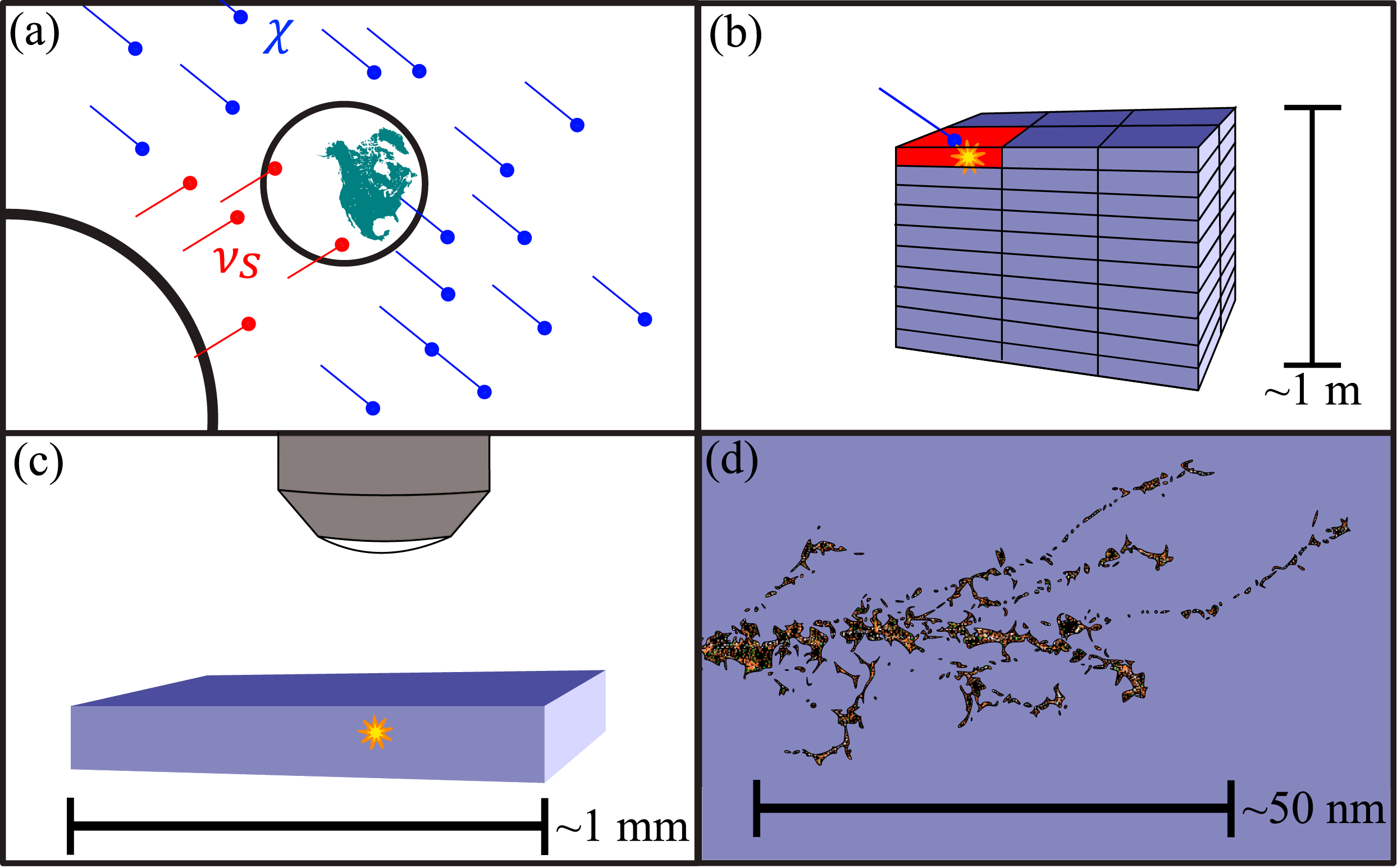}
\end{center}
\caption{\label{fig:DMoverview} Schematic overview of the proposed solid-state directional dark matter detection procedure. {\bf (a)} Solar neutrinos ($\nu_S$) stream towards the earth. Because of the motion of the solar system in the Galactic rest frame, there is a preferred direction of incoming WIMP particles ($\chi$) on the earth (so-called WIMP ``wind''), distinct from solar neutrinos, allowing directional discrimination (see text for details). {\bf (b)} WIMP or neutrino particles interact occasionally with the shielded solid-state detector at an underground facility, depositing energy as well as causing cascades of nuclear recoils, leaving long-lasting crystal damage tracks. Charge, phonon, or photon collection is used to detect and localize the event to a mm-scale segment of the detector. The time of the event determines the orientation of the detector with respect to both the Sun and the galactic WIMP wind. {\bf (c)} The triggered segment is removed from the bulk of the detector. Using optical-diffraction-limited methods applied to quantum defects in the solid, damage is localized in a micron-scale volume. {\bf (d)} The nanoscale structure of crystal damage is mapped using superresolution optical methods or x-ray microscopy, allowing WIMP and neutrino events to be distinguished. From M. C. Marshall, M. J. Turner, M. J. H. Ku, D. F. Phillips, and R. L. Walsworth, Quantum Sci. Technol. 6, 024011 (2021). Reproduced by permission of IOP Publishing Ltd.}
\end{figure}

We envision integrating directional detection within a conventional WIMP detector with well-developed instrumentation and background discrimination methods \cite{Rajendran:2017ynw,Marshall:2020azl}. Hybrid detectors like this would be able to detect candidate WIMP events using established methods for semiconductor solid-state detectors, such as charge, phonon, and scintillation detection  \cite{SuperCDMS_LopezAsamar:2019smu,SuperCDMS_Rau:2020abt,DAMIC:2021crr,DAMIC:2021esz}. To operate at the neutrino floor or below, meter-scale solid-state detectors are needed (see Sec.\,\ref{sec_neutrinofloor_subsection}). Such a detector could time-stamp and coarsely localize events at the mm scale using fabricated charge-collection electrodes or phonon sensors \cite{EDELWEISS_2009_design,SuperCDMSSoudan_2013_design}. When an event is detected in a specific mm-scale chip within a meter-scale, modular detector, that chip can be extracted and interrogated to determine the direction of damage, while the remainder of the detector continues to operate. However, using existing methods, it would be time-prohibitive to scan an entire mm-scale chip with nanoscale resolution to locate and map the orientation of the damage track. We therefore propose a two-step damage track reconstruction strategy within the mm-scale chip: (i) a micron-scale localization of the damage track using optical-diffraction-limited imaging techniques, followed by (ii) high-resolution 3D reconstruction of the damage track. Therefore, the proposed meter-scale detector's readout can be summarized in three steps as follows (see Figure\,\ref{fig:DMoverview}):
\begin{itemize}
    \item {\bf STEP I:} Event detection and localization at the mm scale using charge, phonon, or photon collection. The event time is recorded to determine the absolute orientation of the specific mm-scale chip in which the event occurred.
    \item {\bf STEP II:} Damage track localization at the micron scale using optical-diffraction-limited techniques utilizing quantum defects in the solid. (See sections~\ref{sec_strain_spectroscopy} and \ref{sec_FNTD} for methods.)
    \item {\bf STEP III:} Mapping damage tracks at the nanoscale using either superresolution optical methods or x-ray microscopy. (See sections~\ref{sec_xray_microscopy} and \ref{sec_superresolution} for methods.) The meter-scale detector continues operation during steps II and III.
\end{itemize}

For this hybrid detection method, we propose to develop a detector with diamond as its target material. Diamond is currently being studied for next-generation semiconductor-based detectors because of its excellent semiconductor properties, as well as its high sensitivity to low-mass DM candidates (thanks to the light mass of the carbon nucleus). The implementation of traditional event detection techniques in diamond is an active topic of research \cite{Kurinsky:2019pgb}; this includes the first demonstration of a diamond calorimeter coupled to a transition edge sensor (TES) at cryogenic temperatures \cite{Canonica:2020omq}. Diamond is a promising material for semiconductor electronics applications \cite{diamond_electronics_2002,diamond_electronics_2008} as well as for quantum sensing \cite{NV_review_2014_Schirhagl,NV_review_Levine_2019,NV_review_2020_edmonds,sensitivity_Barry2020} and quantum information processing \cite{NV_QInfo_2006,NV_QInfo_2013} applications that make use of diamond's lattice point defects, which exibit optical prepartion and read-out of long-lived electronic spin states. Such technological demand is supported by modern diamond growth techniques using chemical vapor deposition (CVD) \cite{diamond_electronics_2002,NV_review_2020_edmonds}, which enable repeatable, fast, and low-cost growth of uniform crystals. As a result, diamond has also been widely used in particle physics research \cite{diamond_PP_2016_TARUN,diamond_particle_tracking_2022_snowmass}, including in the ATLAS and CMS detectors at the LHC \cite{diamond_LHC_2015_Oh}. The proposed DM directional detector would require $\rm{\mathcal{O}(1\,m^3)}$ of diamond material, comprised of many smaller, modular segments, to achieve sensitivity below the neutrino floor. This level of production is a realistic extension of developments in diamond growth for quantum sensing over the last decade.  In particular, optimized crystal growth protocols \cite{NV_review_2020_edmonds} reliably achieve high-quality semiconductor properties (allowing sensitive charge and phonon extraction) and uniform, low crystal strain (enabling damage track localization in the proposed detector).

In the remaining sections, we provide an overview of methods available for damage track localization and nanoscale mapping (Sec.\,\ref{sec_methods}) as well as a method for generating injected signals (Sec.\,\ref{sec_ion_implantation}). Here, we present an estimation of the signal strength and background rate in order to evaluate the feasibility of methods based on their sensitivity and speed. SRIM simulations indicate that nuclear recoil-induced damage tracks relevant to WIMP events are ${\rm \mathcal{O}(10-100)\,nm}$ in length and create $\mathcal{O}(50-300)$ lattice vacancies \cite{Rajendran:2017ynw}. It is possible to detect the strain caused by such lattice damage via the effect on the optically-active quantum defects in the crystal lattice. Following ref.\cite{Rajendran:2017ynw}, we assume a fractional measured strain of $\Delta x/x\sim10^{-6}$ at a distance of 30\,nm from a single lattice defect. Furthermore, the strain from each individual point defect in a damage cluster can be used to map the spatial distribution of strain from the entire cluster \cite{strain_estimate_1954}. We benchmark damage localization by a simplified model of the strain signal within a ${\rm \mu m^3}$ volume that contains the damage cluster: for each vacancy, a cylinder of uniform strain $\sim10^{-6}$ with height and radius 30\,nm, while the strain outside the cylinder falls off as $1/r^3$ \cite{Marshall:2020azl}.

We benchmark the damage localization time to not exceed the expected event rate in the detector, so the event direction can be measured as each event is detected before the next occurs. (The localization experiment could, however, be parallelized using multiple setups if necessary.) It is expected that the rate of events from coherent solar neutrino scattering, mainly from ${\rm ^8B}$ solar neutrinos, will be approximately $\mathcal{O}(30)$ per ${\rm ton\times year}$ \cite{spectra_2007,solar_nu_flux_update_2016_Bergstrom}. With a low rate of WIMP events, this would limit the localization time to about 10 days. We consider, however, a conservative three-day target between event registration and damage localization at the micron scale to account for a potential WIMP signal at the neutrino floor, as well as background events \cite{Marshall:2020azl}. Note that we assume a detector with sensitivity below the neutrino floor will have fewer background events than neutrino events. Furthermore, quantum diamond microscopy experiments (see sections \ref{sec_strain_spectroscopy}, \ref{sec_FNTD}, and \ref{sec_superresolution}) are expected to take place on-site at a shielded complex that houses the detector, using infrastructure that is constructed from high-purity materials, similar to what is used in current WIMP detectors \cite{XENON_Hamamatsu_2015_highpurity}. Note that the complementary technique of scanning x-ray diffraction microscopy for nanoscale strain mapping requires access to synchrotron facilities (see Sec.\,\ref{sec_xray_microscopy}). Nevertheless, localization of the WIMP/neutrino signal to a small $\rm{\mu m^3}$ volume before transport and shielding during the transport would eliminate significant background contamination. With a reasonable allocation of synchrotron beam time, scanning the micron-scale volumes of interest should also be feasible.

\section{Nitrogen-vacancy centers in diamond}
\label{sec_nv}

\begin{figure*}[htbp]
\begin{center}
\includegraphics[width=0.7\textwidth]{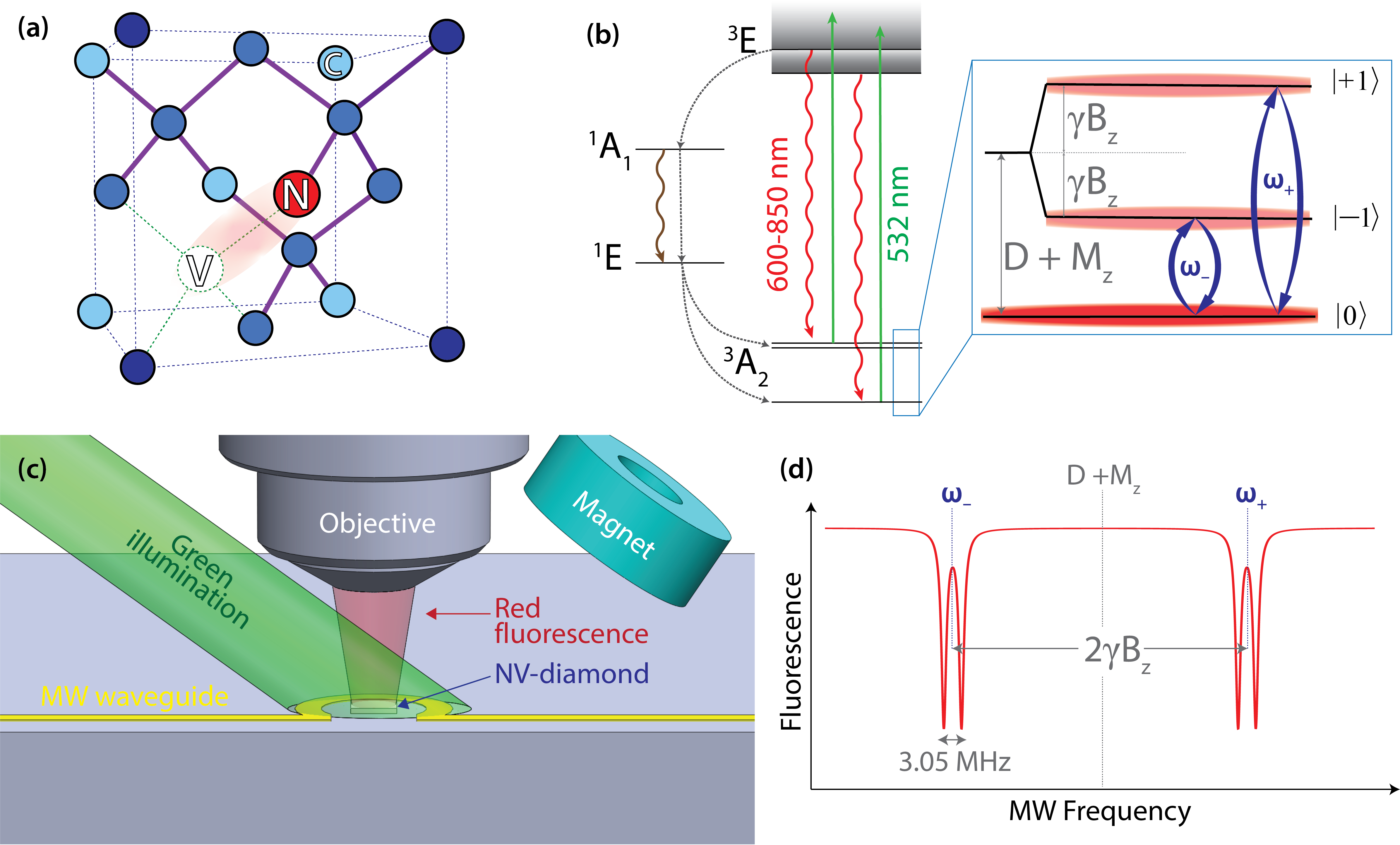}
\end{center}
\caption{\label{fig:NVoverview} Overview of nitrogen-vacancy (NV) centers in diamond. {\bf (a)} Schematic of diamond lattice hosting an NV center. {\bf (b)} NV electronic energy level diagram, illustrating triplet ground and excited state as well as singlet states. Intersystem crossing (ISC) through singlet states yields a nonradiative decay pathway. Inset: Ground state spin energy levels. On excitation by green light, the ISC nonradiative decay path results in dimming of $m_s=\pm1$ states’ red fluorescence relative to that of $m_s=0$. MW frequencies of $\omega_\pm$ can be used to drive spin transitions. Further hyperfine splitting results from electronic spin-nuclear spin interactions (not shown). {\bf (c)} Simplified schematic of a quantum diamond microscope (QDM). Permanent magnets and/or Helmholtz coils (not shown) provide a bias magnetic field to allow spectral resolution of NV spin states via optically detected magnetic resonance (ODMR). Green illumination is used to initialize and read out the spin state. MW delivery via the waveguide can drive spin transitions. A microscope objective collects red fluorescence and focuses it on a camera (not shown). {\bf (d)} Schematic of ODMR data: NV fluorescence as a function of MW frequency. Temperature-dependent zero-field-splitting parameter and local strain determine the center frequency, while the axial bias magnetic field determines the splitting between them. The spin-1/2 nuclear spin of $\rm{^{15}N}$ contributes to the double resonance structure (splitting of about $\rm{3.05\,MHz}$); the $\rm{^{14}N}$ spin-1 nucleus would produce triplet hyperfine energies instead (splitting of about $\rm{2.16\,MHz}$).}
\end{figure*}

Nitrogen-vacancy centers (NVs) in diamond are composed of a nitrogen atom and an adjacent vacancy in a diamond crystal (see Figure\,\ref{fig:NVoverview}a) and can be classified according to their orientation in four crystallographic directions. NV centers exhibit quantized energy levels with optical and spin properties that are similar to atomic systems. The NV energy levels couple to magnetic fields, electric fields, temperature, and strain within the diamond lattice; the resulting effects can be sensitively measured optically, allowing NV centers to function as quantum sensors with good sensitivity and nanoscale spatial resolution under ambient conditions \cite{NVasSensor_2008_Degen,NVasSensor_2008_Taylor,NVasSensor_2008_balasubramanian,NVasSensor_2008_maze,NVasSensor_2008_acosta}. In particular, the NV center is a two-electron system with a triplet spin-1 ground state that typically has a longitudinal relaxation time $T_1\simeq{\rm6\,ms}$ \cite{NVlongitudianlRelaxation_2012_jarmola,NVlongitudianlRelaxation_2014_rosskopf} and decoherence times $T_2$ of up to a few ms \cite{NVdecoherence_2009_balasubramanian} at room temperature.

\textbf{Ground state spin Hamiltonian.} With a bias magnetic field aligned along the NV symmetry axis $z$, the simplified NV electronic ground state Hamiltonian can be written as \cite{Kehayias_2019}
\begin{align}\label{eqn_gsHamiltonian}
    \dfrac{\mathcal{H}_{\rm gs}}{h}\simeq (D+\mathcal{M}_z)S_z^2+\gamma B_zS_z\,.
\end{align}
$D\simeq\rm{2.87\,GHz}$ is the zero-field-splitting (ZFS) parameter at room temperature, resulting from the spin-spin interactions between the two unpaired NV electrons. ZFS varies with temperature as ${\rm d}D/{\rm d}T\simeq{\rm -74.2\,kHz/K}$ \cite{NV_temperatureDependence_2010_acosta}, which is the origin of the NV's use as a nanoscale thermometer. $\mathcal{M}_z$ is the spin-strain coupling parameter, which can be converted to a measure of lattice strain using proper coupling constants  \cite{NVSpinStrainCoupling_2019}. We ignore other strain-coupling parameters since their contribution is subdominant \cite{Kehayias_2019}. The last term is the Zeeman Hamiltonian, which splits the $m_s=\pm1$ eigenstates. The NV electronic gyromagnetic ratio $\gamma=g_e\mu_B/h\simeq\rm{28.03\,GHz/T}$, where $g_e\simeq2.003$ is the NV electronic $g$-factor, $\mu_B$ is the Bohr magneton and $h$ is the Planck's constant. Hamiltonian \eqref{eqn_gsHamiltonian} leads to transition frequencies
\begin{align}
    \omega_\pm\simeq(D+\mathcal{M}_z)\pm\gamma B_z\,.
\end{align}
An energy level diagram of the NV ground state is shown in Figure\,\ref{fig:NVoverview}b (inset). The transitions are in the microwave (MW) frequency range, and can be resonantly driven using MW fields of corresponding frequency.

\textbf{Spin-dependent fluorescence.} NVs can be driven to an excited electronic state by green laser illumination, followed by decay back to the ground state via fluorescence in the red spectrum. In addition, there is a nonradiative decay path for the $m_s=\pm1$ excited state to decay to the $m_s=0$ ground state, which enables spin-state-dependent optical readout and preparation (see Figure\,\ref{fig:NVoverview}b). Resonant MW driving of spin transitions in the NV ground state, in conjunction with the optically-driven dynamics, allows optical determination of the NV spin state through optically detected magnetic resonance (ODMR); see Figure\,\ref{fig:NVoverview}d. Figure\,\ref{fig:NVoverview}c illustrates a simplified schematic of a quantum diamond microscope (QDM), which enables NV-diamond sensing with diffraction-limited resolution.

\begin{figure*}[htbp]
\begin{center}
\includegraphics[width=0.8\textwidth]{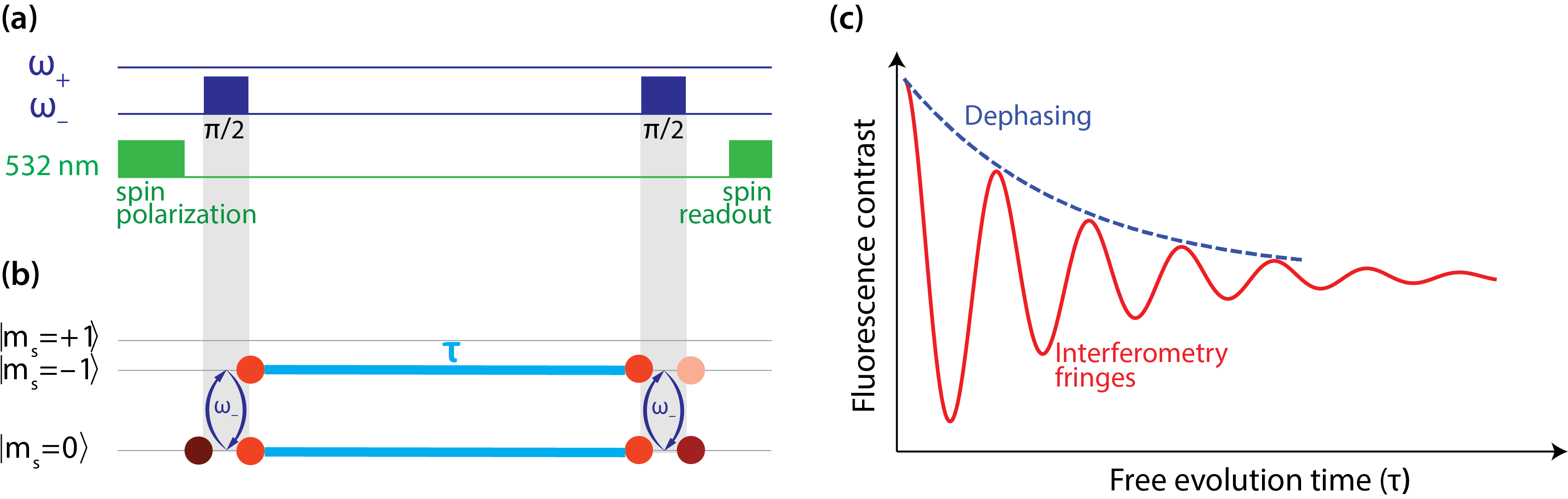}
\end{center}
\caption{\label{fig:Ramsey} Principle of Ramsey measurement protocol in NV systems. {\bf (a)} Ramsey pulse sequence. See text for details. {\bf (b)} Illustration of population in each NV spin state at each point in the Ramey pulse sequence. Following the green initialization pulse, the spin is polarized at $m_s=0$. The first microwave $\pi/2$ pulse induces an equal superposition of $m_s=0$ and $m_s=-1$ states. During free evolution, the two spin state components in the superposition acquire a differential phase due to the effect of terms in the Hamiltonian \eqref{eqn_gsHamiltonian}. The second $\pi/2$ pulse projects the phase onto population difference. Note that in this protocol the $m_s=+1$ state does not play a significant role due to the choice to use microwave pulses resonant with $\omega_-$. {\bf (c)} Schematic NV fluorescence contrast curve as a function of free evolution time. Interferometry fringes reflect the phase accumulated during the free evolution period, which can be $>2\pi$. Inhomogeneous phase evolution causes an overall decay in an ensemble of NV centers on a timescale $T_2^\ast$.}
\end{figure*}

\textbf{Ramsey measurements.} Measurement protocols based on quantum interferometry have been successfully and robustly implemented in NV systems. Here we outline a basic Ramsey protocol to introduce the measurement principle (see Figure\,\ref{fig:Ramsey}); in Sec.\,\ref{sec_strain_spectroscopy}, we present our recent advances in engineered strain-sensitive sensing protocols \cite{marshall_2021_strainCPMG}. Ramsey measurements with NVs use a green laser pulse to initialize the spin state into $m_s=0$. The spin-polarized state is then driven into a coherent superposition state by means of a resonant MW pulse (the so-called $\pi/2$ pulse), which is then allowed to evolve uninterrupted for a period of time. The spin states that comprise the superposition state accumulate a differential phase during the free evolution. Importantly, this differential phase accumulation depends on the transition frequencies encoding the NV coupling to the environment. During the final MW pulse, the evolved NV phase is projected onto a difference in population between $m_s=0$ and $m_s=\pm1$, manifested as interferometry fringes in the measured  NV fluorescence due to the different average fluorescence rates of these spin states (see Figure\,\ref{fig:Ramsey}c). 

\textbf{Ensemble NV-diamond.} Diamonds with ensembles of NV centers provide higher SNR than single NV systems, leveraging statistical averaging over multiple spin sensors (scaling as square root of the number of NVs). Additionally, ensemble of NVs enable widefield imaging via parallel measurement of the spatially distributed NVs using a camera-based detector. High-sensitivity applications using diamond samples with NV concentrations $\rm{\gtrsim ppm}$ ($\rm{10^{17}\,per\,cm^{3}}$) have been demonstrated \cite{NVensembleApp_2015_grezes,NVensembleApp_2015_wolf,NVensembleApp_2016_barry_neuron,NVensembleApp_2018_schloss_vectorMagnetomety,NVensembleApp_2019_zheng}. However, one of the main challenges in using NV ensembles is inhomogeneous dephasing (due to spatial variations in magnetic fields, diamond strain, etc.), leading to a reduced spin dephasing time $T_2^\ast$ (illustrated as the exponentially decaying envelope of the Ramsey curve in Figure\,\ref{fig:Ramsey}c) and broadened ODMR spectral lines. To extend the NV spin dephasing and coherence times, material engineering techniques and dynamical decoupling sequences have been employed, with further improvements envisioned as well \cite{sensitivity_Barry2020}.

\section{Methods for particle directional detection in diamond}
\label{sec_methods}
\subsection{Optical diffraction-limited strain spectroscopy}
\label{sec_strain_spectroscopy}

In this section, we discuss using nitrogen-vacancy (NV) centers for micron-scale localization of WIMP or neutrino-induced lattice damage. Strain features can occur in CVD diamond as localized deformations within the crystal lattice because of imperfections propagated during growth and imperfections in surface processing \cite{strain_interface_origin_2004_martineau,strain_control_2009_friel}. A variety of methods have been developed for measuring and mitigating such strain features in diamond devices \cite{strain_xray_2008_GAUKROGER,strain_control_2009_friel,strain_lowdislocations_typeIIa_2009_martineau_high}. However, these methods rarely meet all the requirements of the directional dark matter detector proposal (see Sec.\,\ref{sec_detector_principle}). The spatial resolution of birefringence imaging is limited since the signal is integrated over the entire diamond thickness \cite{strain_birefringence_2014_hoa,Kehayias_2019}. Raman spectroscopy suffers from a relatively high detection noise floor \cite{strain_raman_1994_alexander,strain_raman_2011_crisci}. X-ray tomography and microscopy involve lengthy data acquisition times as well as access to synchrotron beamlines \cite{strain_xray_imaging_2009_Moore,marshall_2021_xray}. An alternative method is, however, quantum-point-defect strain spectroscopy, which offers both the speed and sensitivity required for Step II of the directional dark matter detection procedure, as outlined above. NV ground state spin transitions are sensitive to local strain (see Sec.\,\ref{sec_nv}); therefore, spectroscopic measurements can be used to detect local strain features using NV centers as integrated quantum sensors \cite{Broadway_2019,Kehayias_2019}. Recent developments in NV strain spectroscopy suggest promising prospects for diamond-based directional dark matter detectors \cite{marshall_2021_strainCPMG}.

\begin{figure*}[htbp]
\begin{center}
\includegraphics[width=0.8\textwidth]{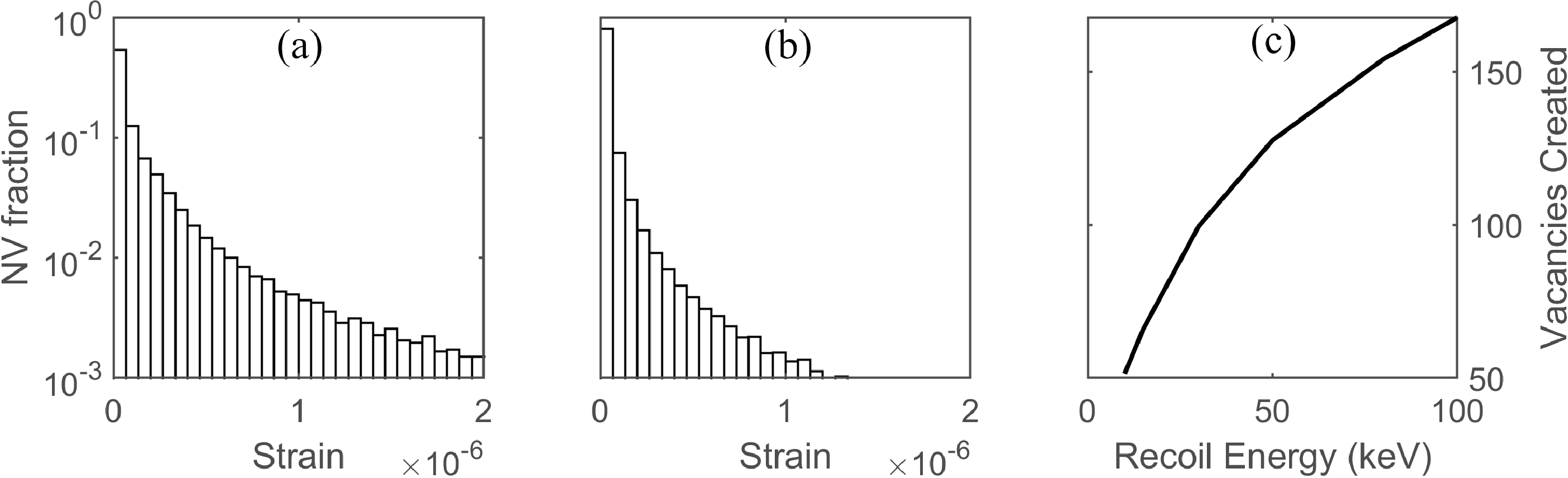}
\end{center}
\caption{\label{fig:strain_dist_SRIM} Calculated fraction of NV centers experiencing a given strain, within {\bf (a)} $\rm{0.1\,\mu m^3}$ volume and {\bf (b)} $\rm{1\,\mu m^3}$ volume from a $\rm{10\,keV}$ nuclear recoil. {\bf (c)} Number of vacancies created as a function of the recoil energy, as predicted by SRIM simulations \cite{SRIM}. From M. C. Marshall, M. J. Turner, M. J. H. Ku, D. F. Phillips, and R. L. Walsworth, Quantum Sci. Technol. 6, 024011 (2021). Reproduced by permission of IOP Publishing Ltd.}
\end{figure*}

As discussed previously \cite{Marshall:2020azl}, we model the strain induced by each WIMP or neutrino-induced vacancy in the lattice as a cylinder of uniform strain $10^{-6}$ with height and radius $\rm{30\,nm}$, while the strain outside the cylinder falls off as $1/r^3$ (see Sec.\,\ref{sec_detector_principle}). Based on this model, Figure\,\ref{fig:strain_dist_SRIM} shows the fraction of NV centers experiencing a given stain for a $\rm{10\,keV}$ recoil, which results in about 50 induced vacancies (Figure\,\ref{fig:strain_dist_SRIM}c). We consider two scenarios for damage track localization: a resolved volume of about $\rm{0.1\,\mu m^3}$ with diffraction-limited $\rm{350\,nm}$ lateral resolution and $1\,\mu m$ axial resolution (Figure\,\ref{fig:strain_dist_SRIM}a); and a resolved volume of $\rm{1\,\mu m^3}$ with three-dimensional $\rm{\mu m}$ resolution (Figure\,\ref{fig:strain_dist_SRIM}b).  NV centers are most sensitive to strain projections onto their symmetry axes. Depending on the relative orientation and location of crystal damage and point defects within the resolved voxel, the strains projected onto different NV centers could add constructively or average to zero. In the former case, we compute the mean averaged strain (leading to frequency shifts), and in the latter case, we calculate the standard deviation of the strain distribution (leading to line broadening and a reduction in NV spin dephasing time). A small number of NVs exhibit considerably higher strain levels $>10^{-5}$, whose effect on the signal depends on experimental details. Combining different cases, the voxel-averaged strain signal ranges between $1\times10^{-7}$ and $3\times10^{-6}$ \cite{Marshall:2020azl}.

\begin{figure*}[htbp]
\begin{center}
\includegraphics[width=0.8\textwidth]{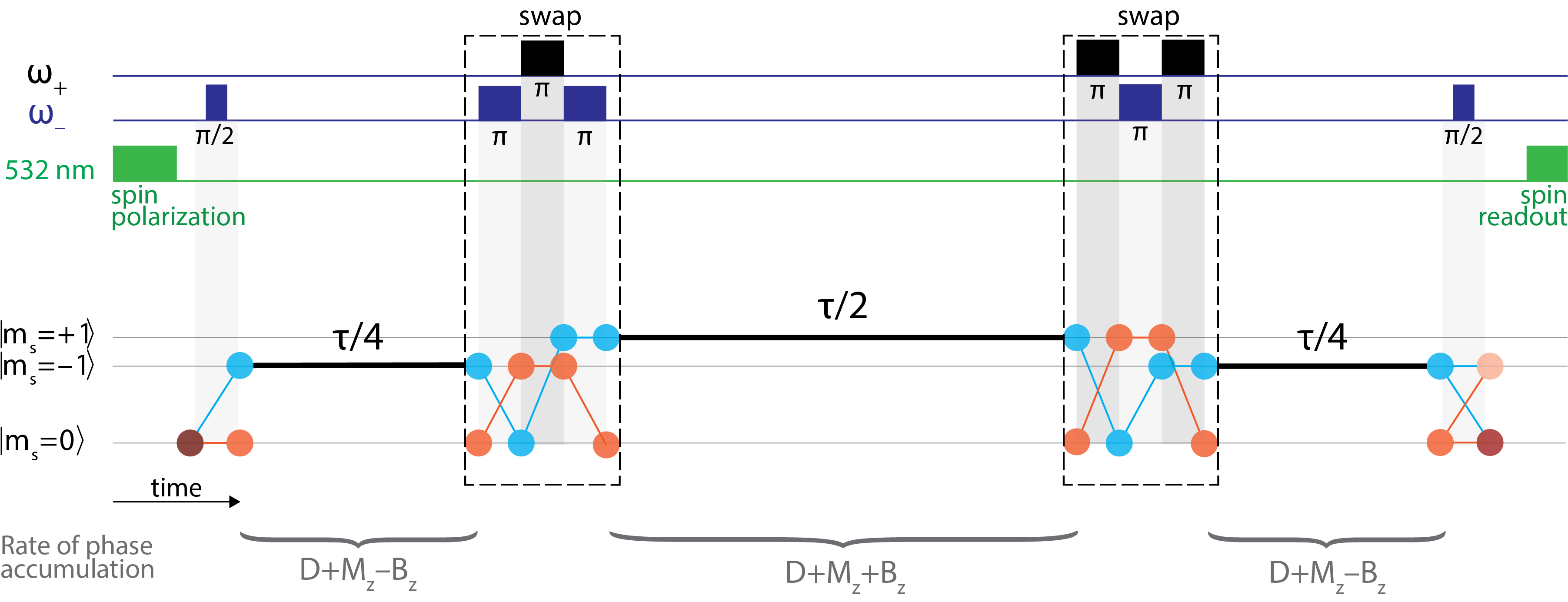}
\end{center}
\caption{\label{fig:strainCPMG} Strain-CPMG measurement protocol. Top: Pulse sequence. $\omega_{\pm}$ denotes microwave frequencies addressing $\rm{\ket{m_s=0}}\leftrightarrow\ket{m_s=\pm1}$ transition. Bottom: Evolution of the ground state spin population. See ref.\,\cite{marshall_2021_strainCPMG} for details.}
\end{figure*}

NV strain imaging is typically performed using a quantum diamond microscope (QDM) \cite{NV_review_Levine_2019} to detect strain-induced modulations in optically detected magnetic resonance (ODMR) spectra \cite{NV_review_Levine_2019,Broadway_2019,Kehayias_2019} (see Figure\,\ref{fig:NVoverview}c and \ref{fig:NVoverview}d). However, the limited sensitivity of this approach would require lengthy averaging times (many days) to detect WIMP-induced strain and hence render it unsuitable for dark matter detection. The implementation of Ramsey-like measurements in a QDM, however, provides fast and sensitive measurements ($\sim\,$few hours). The strain Carr-Purcell-Meiboom-Gill (strain-CPMG) sequence, illustrated in Figure\,\ref{fig:strainCPMG}, is a Ramsey-like sequence that is suitable for strain sensing \cite{marshall_2021_strainCPMG} (and thermometry in certain applications \cite{Thermometry_2013_Lukin,Thermometry_2013_Awschalom,Thermometry_2013_Wrachtrup,Thermometry_2015_Wang,DRamsey_2018_KonzelmannCoop}). The protocol is insensitive to magnetic inhomogeneities arising from the electronic and nuclear spin bath within the diamond lattice, allowing longer NV ensemble spin dephasing times and hence enhanced strain sensitivity. In particular, by exchanging population between $\ket{\rm{m_s=-1}}$ and $\ket{\rm{m_s=+1}}$ levels (which have opposite dependence on magnetic field) during the free evolution time of a Ramsey sequence, the magnetic-field-induced precession cancels (Figure\,\ref{fig:strainCPMG}) and the strain sensitivity is greatly enhanced \cite{marshall_2021_strainCPMG}.

We describe details of the spin evolution during the strain-CPMG protocol here. We first spin polarize the NV ground state to $\ket{{\rm m_s} = 0}$. After that, we prepare an equal superposition of $\ket{0}$ and $\ket{-1}$ by applying a microwave (MW) $\pi/2$ pulse resonant with the $\ket{0}\leftrightarrow\ket{-1}$ transition. The NV spins accumulate a relative phase with the rate of $D+M_z-\gamma B_z$ (see eq.\,\eqref{eqn_gsHamiltonian}) during the free evolution of this superposition state. Following that, we apply triplets of MW $\pi$-pulses, switching between the $\ket{0}\leftrightarrow\ket{-1}$ and $\ket{0}\leftrightarrow\ket{+1}$ transitions in order to collectively swap the NVs from $\ket{-1}$ to $\ket{+1}$. As a result, during the middle free evolution time $\tau/2$ the phase accumulation rate is $D + M_z + \gamma B_z$. We next apply a reverse swap pulse to transfer the spin population from $\ket{+1}$ back to $\ket{-1}$ and phase accumulation continues for another $\tau/4$ duration. The final accumulated NV phase is independent of $B_z$ and only sensitive to $D+M_z$. The last MW $\pi/2$ pulse projects the accumulated phase onto the NV population difference in the $\ket{0}$ and $\ket{-1}$ states, which we read out using the spin-dependent fluorescence measurement discussed in Sec.\,\ref{sec_nv}.

The strain-CPMG protocol employed on a QDM enabled us to perform high-precision strain mapping at micron-scale resolution and with mm-scale field-of-view \cite{marshall_2021_strainCPMG}. The technical details of this work are presented in ref.\,\cite{marshall_2021_strainCPMG}, but here we summarize the most important implications for directional dark matter detection. We use a low-strain CVD bulk diamond grown by Element Six Ltd. with  NV concentrations of about 0.4\,ppm. The results obtained with this diamond sample are a promising example of state-of-the-art diamond growth technology relevant to the dark matter detection proposal. This sample is isotopically purified with 99.995\% $\rm{^{12}C}$; however, we expect that the strain-CPMG sequence will not be significantly affected by higher $\rm{^{13}C}$ spin concentration with nuclear spin $I=1/2$ (this reduced requirement might simplify the crystal growth process for scaled-up dark matter detectors). In particular, our strain-CPMG measurements exhibit about three times longer dephasing times than the basic Ramsey protocol, showing strain-CPMG's ability to reduce the effects of inhomogeneous magnetic noise.

\begin{figure*}[htbp]
\begin{center}
\includegraphics[width=0.8\textwidth]{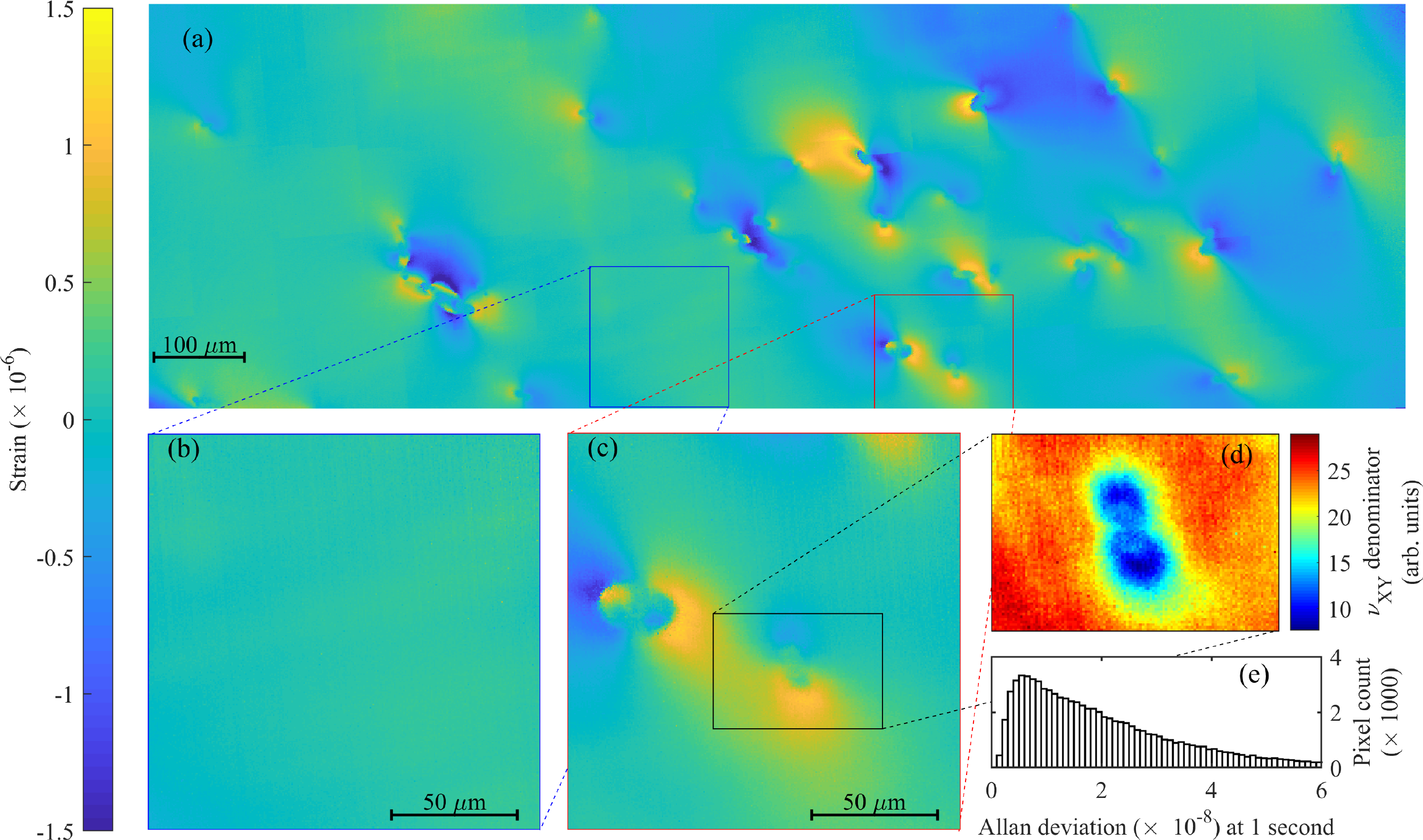}
\end{center}
\caption{\label{fig:imagerone} 2D strain map of a diamond chip generated with a quantum diamond microscope (QDM) using the strain-CPMG measurement protocol. {\bf(a)} Manually-registered array of multiple $\rm{150\times150\,\mu m^2}$ strain images, covering a mm-scale area of the NV-diamond sample, with a 2D pixel size of about $\rm{0.5\,\mu m^2}$. Each field-of-view is acquired in one second. Since diamond is an excellent thermal conductor \cite{thermal_conductivity_2001_Twitchen}, the temperature is constant over the entire sample area for a single measurement, however, temperature drift between multiple measurements causes minor artifacts in the concatenated image. {\bf(b)} An example of a low-strain region. {\bf(c)} Typical intrinsic strain features are detected. {\bf(d)} Interferometry curve amplitude (refer to ref.\cite{marshall_2021_strainCPMG} for definition of the visibility parameter $\nu_{XY}$) in a high-strain-gradient region with degraded dephasing time. {\bf(e)}  Distribution of the pixel Allan deviation after one second of averaging data in the low-strain region (b). Reprinted with permission from M. C. Marshall, R. Ebadi, C. Hart, M. J. Turner, M. J. H. Ku, D. F. Phillips, and R. L. Walsworth, Phys. Rev. Applied 17, 024041 (2022). Copyright (2022) by the American Physical Society.}
\end{figure*}

In the first step, we employ a confocal microscope to limit the interrogation volume to micron scale, and characterize the volume-normalized strain sensitivity of the method; we obtain an unprecedented value of $\rm{5(2)\times10^{-8}/\sqrt{Hz\cdot\mu m^{-3}}}$, surpassing previous work \cite{Kehayias_2019} by two orders of magnitude. To map mm-scale areas, we use a widefield-imaging QDM employing the strain-CPMG protocol, because confocal scanning is excessively time-consuming. Figure\,\ref{fig:imagerone} shows the widefield imaging results. A mm-scale strain map is obtained by registering multiple fields-of-view of $\rm{150\times150\,\mu m^2}$ each with one second of data acquisition (Figure\,\ref{fig:imagerone}a). We detect strain features of strength $\sim10^{-6}$ at micron-scale. Large areas of diamond sample with sub-$10^{-7}$ strain variations are observed, indicating promising prospects for diamonds used to detect dark matter. An example of such a low-strain region is shown in Figure\,\ref{fig:imagerone}b, and a distribution of one-second pixel Allan deviations is shown in Figure\,\ref{fig:imagerone}e. Furthermore, post processing might be performed in order to improve spatially resolving WIMP/neutrino-induced strains; intrinsic strain features can be spatially filtered, for instance, using high-pass filters or with modern machine learning-based methods proposed in similar contexts \cite{emulsions_ML_2021}. Because of the fast decay of the signal, interferometric measurements are less sensitive in pixels with large strain variations. These pixels, for example, can be found near the center or edge of strain features. We can measure the amplitude of the interferometry curve using the two quadratures of the interferometry signal (see ref.\cite{marshall_2021_strainCPMG} for details). Figure\,\ref{fig:imagerone}d shows an example of how we use this information to identify pixels with high strain gradients.

Since the QDM used for measurements reported in Figure\,\ref{fig:imagerone} does not impose any further depth restrictions beyond the focal plane of the objective, sufficient z-resolution for 3D micron-scale localization remains to be demonstrated. Even high numerical aperture optics, which restrict depth of field in principle, collect out-of-focus light, especially given the high refractive index of diamond. However, high axial resolution can be achieved using optical sectioning methods which remove out-of-focus light contributions \footnote{An alternative approach towards higher axial resolution is z-dependent NV control with engineered spatial gradients of the MW drive \cite{inhomogeneous_MW_eng_2019}.}. Structured illumination microscopy (SIM) \cite{Saxena:2015,Chasles:2006} and light-sheet microscopy (LSM) \cite{Olarte:2018,Keller:2015} are two well-established optical sectioning methods that have numerous applications, particularly in the life sciences. These techniques can be adapted to NV-diamond microscopy. 

In the SIM method, a known spatially varying illumination pattern is created via introducing an optical grating. After acquiring a set of images with different phases of the illumination pattern, post-processing removes the superimposed pattern, as well as out-of-focus contributions encoded in modulations of the introduced pattern. SIM is a wide-field technique, which is crucial for the fast micron-scale localization in the proposed directional WIMP detection scheme. It also has the advantage of simple technical implementation that is compatible with other microscopy techniques \cite{Saxena:2015}. However, SIM typically involves illumination patterns that are not uniform in all directions, which leads to direction-dependent artifacts in the reconstructed images. Such artifacts can be reduced (or vetoed) by applying the pattern at different angles or using different patterns \cite{Yang:2014}. SIM has been used to achieve superresolution imaging of near-surface NV patterns in a bulk diamond \cite{Yang:2014}, however, implementation of SIM to achieve $\mu{\rm m}$ z-resolution remains to be demonstrated. 

In the LSM method, a sheet of light is passed through the sample and the induced fluorescence is collected through a separate objective. In most cases, Gaussian beams and cylindrical lenses are used to generate the light sheet. The light sheet and collection axis are typically perpendicular, with the collection objective aligned so that its focal volume matches the light sheet.  Fast and high-resolution 3D microscopy can be achieved using LSM. Although LSM has its advantages, there are also some challenges: e.g., broadening or attenuation of light sheets, scattering and absorption of fluorescence light generated deep within a sample, sample-induced aberrations, and well-known stripe artifacts resulting from absorption and scattering from the illumination side of the sample \cite{Olarte:2018}. It is possible to mitigate some of the mentioned effects in post-processing or even during image acquisition through modified LSM implementations \cite{Olarte:2018}. A potential approach to counteract sample-induced aberration is to use adaptive optics tools, which have shown utility in a variety of applications, such as astronomical telescopes \cite{Beckers:1993} and biomedical imaging \cite{Godara:2010}. LSM has been implemented in an NV-diamond experiment to characterize microwave devices \cite{Horsley:2018}, although the light sheet thickness of 14 $\mu{\rm m}$ does not yet meet requirements for dark matter detection. LSM is a rapidly advancing technique, particularly in the life sciences. Achieving diamond strain spectroscopy with three-dimensional um-scale resolution will require adaptation of state-of-the-art LSM methods to NV-diamond microscopy.

While 3D resolution $\sim\rm{\mu m^3}$ has not been demonstrated, the volume-normalized sensitivity measured by a confocal scanning QDM shows that detection of WIMP-induced strain features should be possible. Assuming the reported time and volume-normalized sensitivities, the minimum estimated voxel-averaged strain signal ($\sim10^{-7}$) could be detected in a $\rm{mm^3}$ diamond within 9 hours using a widefield QDM setup. Additionally, the technical simplicity and relatively low cost of QDMs makes it possible to use multiple setups simultaneously, e.g., to probe different sections of a mm-scale diamond chip; or several mm-scale chips in parallel, if needed, depending on the rate of neutrino and WIMP interaction. Thus, the localization time is well within the benchmark of three days.

\subsection{Fluorescence detection of defect creation}
\label{sec_FNTD}

Under high-temperature annealing, diamond lattice vacancies become mobile and can be captured by fixed nitrogen impurities, forming nitrogen-vacancy (NV) centers \cite{NVFormation_2014_Theory,NVFormation_simulation_2014,NVFormation_2017_Haque,NVFormation_kinetics_2020}. This NV center creation mechanism can be utilized as an alternative approach to localize a neutrino or WIMP-induced damage track \cite{Marshall:2020azl}: a scattering event induces lattice vacancies along with interstitial nuclei;  subsequently, high-temperature treatment of the sample helps develop NV centers preferentially at the damage track, which can then be detected through fluorescence microscopy. Fluorescent nuclear track detection using $\rm{Al_2O_3:C,Mg}$ crystals \cite{FNTD_original_2006} is a mature field with applications in oncology/dosimetry \cite{FNTD_HybridCell_2013,FNTD_AlphaDosimetry_2018}, as well as nuclear \cite{FNTD_NuclearCharge_2014}, neutron \cite{FNTD_Neutrons_2018}, and beam \cite{FNTD_Simulation_2017} physics. Such fluorescent nuclear track detectors (FNTDs) capture particle tracks in three dimensions through the electronic activation of color centers in the crystal lattice caused by the local electron cascades triggered by particle interactions \cite{FNTD_3D_2013}. However, FNTDs are focused on much longer track lengths than the expected damage length of WIMP events. Recently, low energy nuclear recoil detectors employing color center creation in crystal targets have also been proposed, with potential application as neutrino and light dark matter detectors \cite{colorcenter_Budnik:2017sbu,colorCenters_Cogswell:2021qlq,snowmass_colorCenters}.  Fluorescence detection of particle tracks in diamond has been demonstrated \cite{Diamond_FNTD_2015}; for example, to study diffusion dynamics of the particle-induced vacancies \cite{Diamond_FNTD_2017_V_diffusion}.  Nonetheless, the ability to detect lower energy particles relevant for WIMP recoils remains to be established.

\begin{figure}[htbp]
\begin{center}
\includegraphics[width=0.48\textwidth]{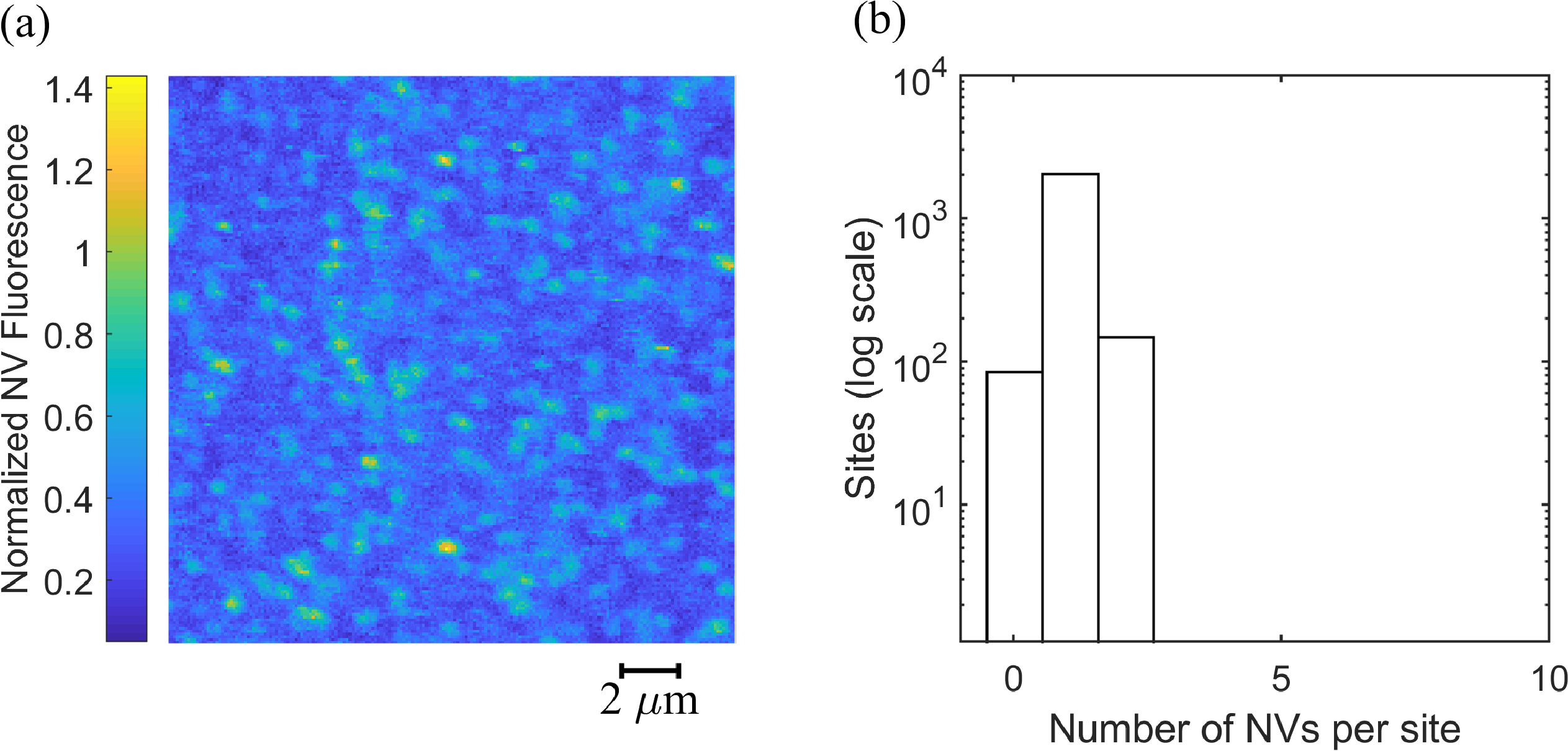}
\end{center}
\caption{\label{fig:hpht_background} Nitrogen-vacancy center background in type Ib HPHT diamonds. The measurements are carried out using a scanning confocal microscope. The number of NVs is determined based on the normalized intensity of fluorescence at each site. {\bf (a)} Typical field of view illustrating intrinsic NV centers. {\bf (b)} Distribution of pre-existing NV centers in diffraction-limited spots. Approximately $\rm{10^4\,\mu m^3}$ volume of diamond from three different samples is scanned. A maximum of two NVs per spot is found. Note that the low background facilitates fast confocal scanning; with about 100 $\rm{\mu s}$ of dwell time, a mm-scale diamond can be scanned within one day. From M. C. Marshall, M. J. Turner, M. J. H. Ku, D. F. Phillips, and R. L. Walsworth, Quantum Sci. Technol. 6, 024011 (2021). Reproduced by permission of IOP Publishing Ltd.}
\end{figure}

Diamonds of low pre-existing NV concentration and high nitrogen impurity content offer favorable prospects of localizing WIMP damage tracks using an NV creation scheme. A high concentration of pre-existing NV centers would present a large background for WIMP-induced NVs. HPHT-grown type Ib diamonds typically contain few vacancies and pre-existing NV centers, while nitrogen concentration can be as high as a few hundred ppm \cite{N_NV_HPHT_2018,NV_HPHT_CVD_TEM_2017}. The initial scans reported in ref.\,\cite{Marshall:2020azl} find no more than two NV centers in the diffraction-limited volume of such a diamond sample (see Figure\,\ref{fig:hpht_background}). Assuming similar NV background in a scaled up diamond detector material, three or more NVs created for each WIMP event would be sufficient for diffraction-limited damage track localization. Simulations suggest that each WIMP scattering event results in $\mathcal{O}(10-100)$ lattice vacancies (see Figure\,\ref{fig:strain_dist_SRIM}c). However, not all the vacancies form NV centers during high-temperature annealing -- competing processes like recombination of a vacancy with an interstitial carbon atom or formation of vacancy clusters reduce the NV yield \cite{NVFormation_2014_Theory,NVFormation_simulation_2014,NVFormation_2017_Haque}. In order to determine the minimum detectable recoil energy, NV yield must be characterized for different incident particle energies and annealing parameters. An understanding of NV creation efficiency can be gained from molecular dynamics simulations of induced damage and annealing processes \cite{NVFormation_simulation_2014}. Additionally, the described characterization can be performed experimentally employing focused ion beam (FIB) systems (see Sec.\,\ref{sec_ion_implantation}).

If damage localization through NV creation proves to be effective, a crucial aspect of this method that needs to be demonstrated is maintaining the directionality and head/tail asymmetry of the damage tracks (see Figure\,\ref{fig:SRIM}) during high-temperature annealing. For the 3D directional information to be retained, vacancy travel range must be less than 10 nanometers since damage track length scale is expected to be several tens of nanometers. Twenty nitrogen atoms are located within a 3 nm travel range in a type Ib diamond with 200 ppm nitrogen concentration; the expected vacancy capture probability of $>1\%$ \cite{NVFormation_simulation_2014} suggests minor damage geometry deformation. The amount of directional signal wash-out caused by annealing can be assessed by nanoscale analysis of the damage track induced by single ion implantation before and after annealing. Note that the high nitrogen content in type Ib diamonds may help preserve the head/tail asymmetry, but it reduces the NV coherence time, which negatively affects nanoscale mapping using conventional superresolution Ramsey-type measurements (see Sec.\,\ref{sec_superresolution}). The strain-CMPG protocol can be employed to extend the coherence time by mitigating magnetic noise of the nitrogen content. Alternatively, the directionality readout could be achieved by resolving the positions of individual NVs or using x-ray diffraction microscopy \cite{marshall_2021_xray}. 

\subsection{X-ray diffraction microscopy}
\label{sec_xray_microscopy}

Scanning x-ray diffraction microscopy (SXDM) offers a method to map crystal strain features in 3D at the nanoscale \cite{Marshall:2020azl}. In a recent study \cite{marshall_2021_xray}, using the Hard x-ray Nanoprobe (HXN) at Argonne National Laboratory's Advanced Photon Source \cite{APSNanoprobe2012}, we demonstrated SXDM's capabilities for reconstructing in 3D the expected strain induced in diamond by a WIMP or neutrino creating a damage track. HXN uses a monochromatic beam of hard x-rays focused to a spot size of $\rm{10-25\,nm}$ \cite{HoltReview2013,MohacsiFresnelOptics2017}. A pixelated photon counter records the spatial pattern of diffracted photons as the spot moves across the sample. These diffraction patterns encode information about local crystallographic deformations \cite{xray_book_Warren_1990}. The spot size resolution of about 10 nm is expected to be sufficient for directional signal determination, but even higher resolutions can be achieved, for example, using Bragg projection ptychography \cite{xrayptychography_Pfeiffer_2018,xrayptychography_3DBPP2017,xrayptychography_multiangleBPP2018}.

\begin{figure}[htbp]
\begin{center}
\includegraphics[width=0.35\textwidth]{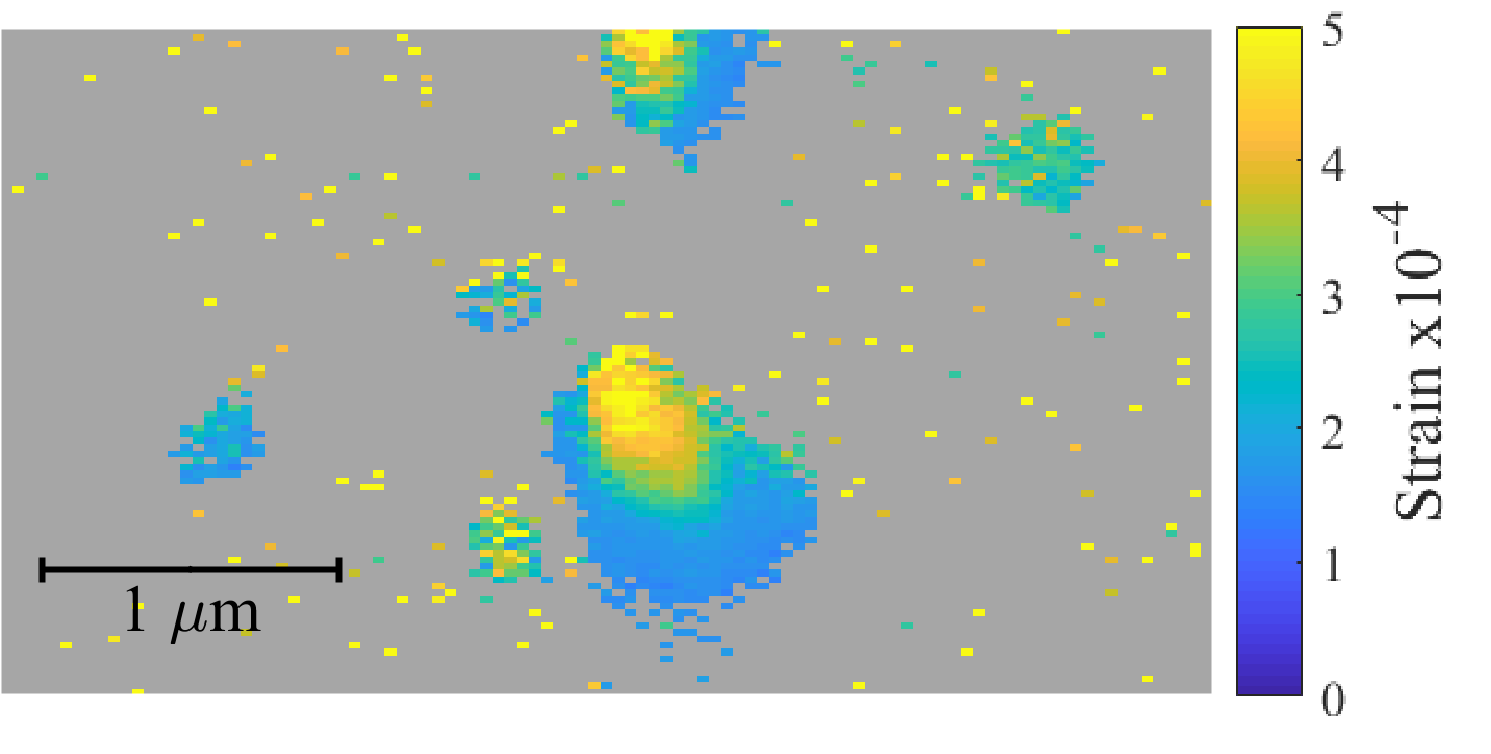}
\end{center}
\caption{\label{fig:xray_sage_HPHT} SXDM map of strain features in HPHT diamond at a length scale similar to that expected for WIMP signals are detected using SXDM. The results demonstrate SXDM's ability to resolve damage caused by WIMP or neutrino in diamond crystals. Scanning step size is 22 nm. Reprinted with permission from M. C. Marshall, D. F. Phillips, M. J. Turner, M. J. H. Ku, T. Zhou, N. Delegan, F. J. Heremans, M. V. Holt, and R. L. Walsworth, Phys. Rev. Applied 16, 054032 (2021). Copyright (2021) by the American Physical Society.}
\end{figure}

\begin{figure}[htbp]
\begin{center}
\includegraphics[width=0.35\textwidth]{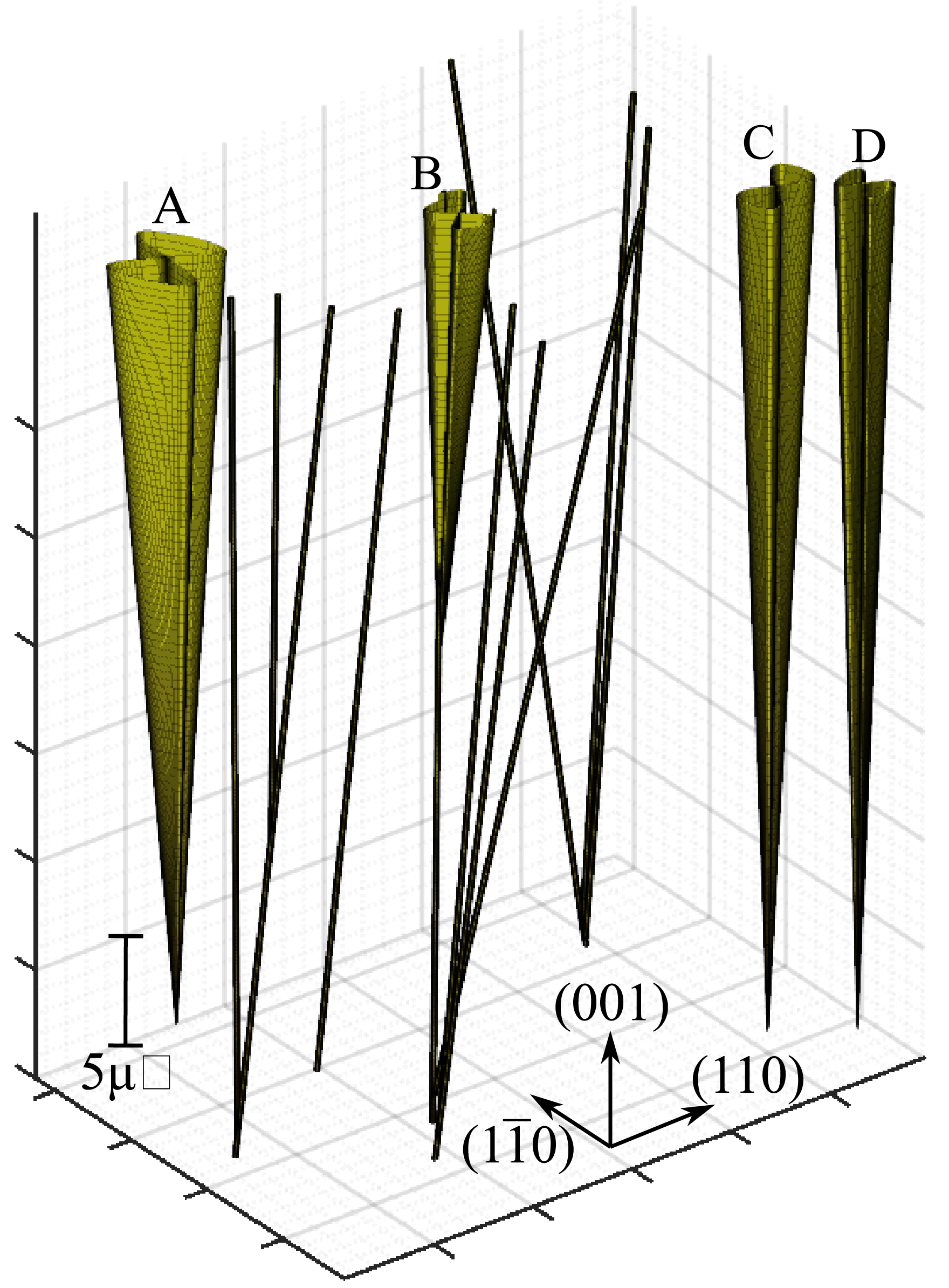}
\end{center}
\caption{\label{fig:xray_3d} Reconstruction of the intrinsic strain features in a CVD diamond in three dimensions. 3D models are constructed using SXDM scans in the $(113)$ and $(\Bar{1}13)$ diffraction planes. The illustrated boundary of features correspond to strain $2\times10^{-4}$. ``Rodlike'' features are likely the high-strain edges of similar dislocation features with small projections onto both of the diffracting planes. See ref.\,\cite{marshall_2021_xray} for details. Reprinted with permission from M. C. Marshall, D. F. Phillips, M. J. Turner, M. J. H. Ku, T. Zhou, N. Delegan, F. J. Heremans, M. V. Holt, and R. L. Walsworth, Phys. Rev. Applied 16, 054032 (2021). Copyright (2021) by the American Physical Society.}
\end{figure}

Figure\,\ref{fig:xray_sage_HPHT} shows a high resolution ($\rm{\sim20\,nm}$) scan of strain features in an HPHT diamond sample, demonstrating the ability to resolve features with sufficient spatial and strain resolution to detect WIMP or neutrino-induced damage tracks. Using the strain model associated with WIMP-induced vacancies (see Sec.\,\ref{sec_detector_principle}), the expected damage strain from a 10 keV nuclear recoil is $1.8\times10^{-4}$ at a 30 nm spot. Ref.\,\cite{marshall_2021_xray} reports a strain sensitivity of about $1.6\times10^{-4}$. Therefore, the WIMP signal should be detectable. The CVD diamond sample used for this sensitivity analysis includes multi-layer strain structures, which complicate the analysis and possibly reduce the sensitivity. Uniform diamond samples with low strain, ideal for detecting dark matter, can result in up to about an order of magnitude lower strain detection floor. In addition, scanning the crystal from different Bragg angles allows for 3D reconstruction of strain features. In Figure\,\ref{fig:xray_3d}, intrinsic growth strain features are scanned at two different Bragg angles and 3D reconstructed using a model described in ref.\cite{marshall_2021_xray}. In summary, SXDM proves the ability to detect directional damage in diamond by combining high spatial resolution, high strain resolution, and 3D reconstruction capabilities. 

\begin{figure}[htbp]
\begin{center}
\includegraphics[width=0.48\textwidth]{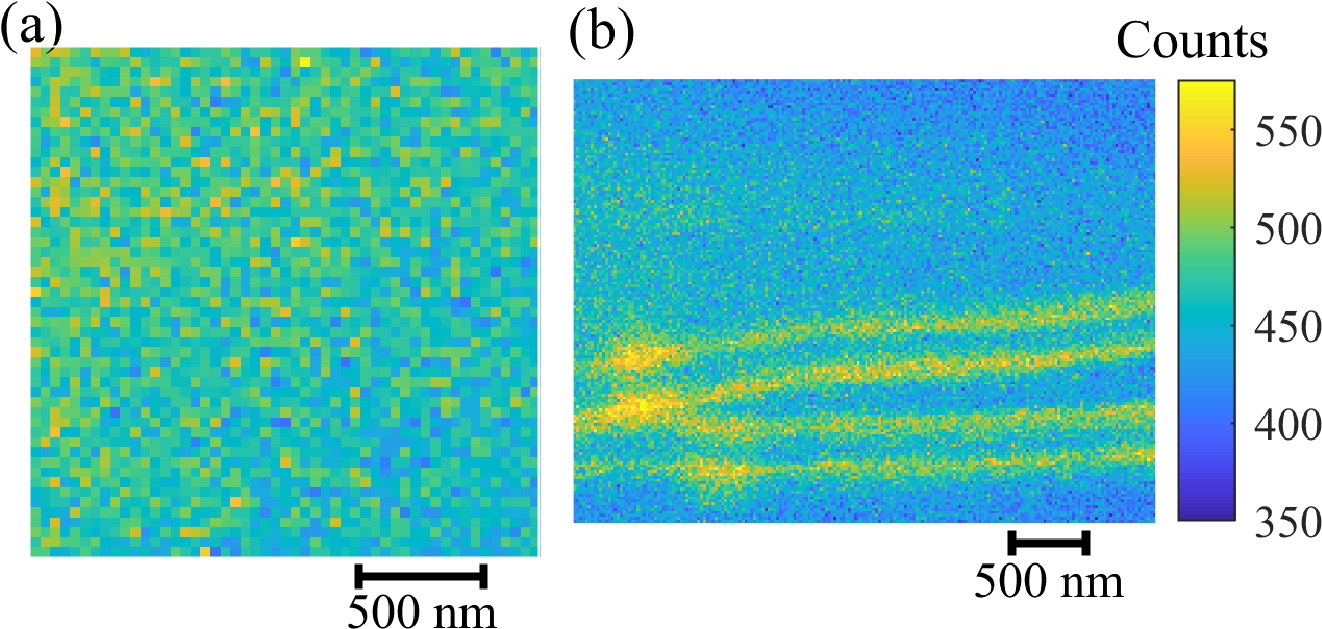}
\end{center}
\caption{\label{fig:xray_backround} In a CVD diamond, small-scale background SXDM scans are performed far away from large-scale strain features. Results demonstrate no preexisting strain tracks corresponding to the expected neutrino or WIMP-induced strain features, i.e., localized strain features at $\rm{\lesssim100\,nm}$. Plots show the number of detector counts attributable to strain in the CVD diamond layer. A scanning step size of {\bf (a)} 40 nm and {\bf (b)} 20 nm was used for two different diamond regions. Reprinted with permission from M. C. Marshall, D. F. Phillips, M. J. Turner, M. J. H. Ku, T. Zhou, N. Delegan, F. J. Heremans, M. V. Holt, and R. L. Walsworth, Phys. Rev. Applied 16, 054032 (2021). Copyright (2021) by the American Physical Society.}
\end{figure}

Preexisting localized sub-micron strain features can be identified as ``false positive'' dark matter signals. A nanoscale study of such intrinsic small-scale strain features in CVD diamond samples is crucial for the realization of diamond-based detectors. Figure\,\ref{fig:xray_backround} illustrates an initial background search in a limited scanning region. In this survey, we do not find any false positives. We do not see any localized sub-micron features in Figure\,\ref{fig:xray_backround}a. We detect features in Figure\,\ref{fig:xray_backround}b that are 100 nm wide, but they extend over a few microns in one dimension, which is inconsistent with WIMP or neutrino-induced damage tracks. Even though further background characterization is still needed, the results suggest that defect patterns in CVD diamond are either pointlike or extended, thus, distinguishable from WIMP or neutrino signals. The observed defect spatial structure is also consistent with those expected of CVD samples treated by high-temperature annealing \cite{PintoExtendedDefects,strain_control_2009_friel}.

At present, only a few synchrotron facilities worldwide offer hard x-ray nanobeams with suitable resolution and sensitivity \cite{APSNanoprobe2012,xray_hxn_nsls,xray_nanomax_sweden}, and high-resolution scans are rather time-consuming. However, with accurate optical localization for candidate WIMP events, the damage track direction for all WIMP events could be determined with a reasonable amount of x-ray beam time.  Furthermore, the detector segments would need to be adequately shielded against cosmic rays in transit from the underground detector site to the synchrotron facility. (That being said, we note that the shielding in transit can be less extensive than that in the underground facility with the $\mathcal{O}({\rm m^3})$ full detector, because the expected event rate in a small $\rm{\mu m^3}$ volume of interest is much lower.) 

\subsection{Superresolution NV microscopy}
\label{sec_superresolution}

In addition to hard x-ray diffraction microscopy, superresolution techniques can also be used to resolve directional WIMP tracks below the optical diffraction limit \cite{Rajendran:2017ynw,Marshall:2020azl}. We summarize the current state-of-the-art in the field and outline a path toward three-dimensional reconstruction of WIMP tracks.

Superresolution techniques are mainly being applied to NV-diamond systems in the context of nanoscale magnetic resonance imaging \cite{NV_review_2014_Schirhagl,NVMRIreview2016,NVMRIreview2019}. By using relevant measurement protocols, such as strain-CPMG, these techniques can be adapted for strain sensing. One class of techniques uses a secondary doughnut-shaped optical beam (created via a helical waveplate), superimposed with the Gaussian-shaped excitation beam, to suppress NV fluorescence signals from everywhere except the low-intensity central region of the doughnut beam. The nonlinear nature of such suppression allows location of the NV emitter to much below the diffraction limit, down to about 10 nm. For example, STED \cite{STED_Early_Hell2009,STED_SIL2012,STED_nanodiamonds2013} causes stimulated emission depletion under the doughnut; CSD \cite{CSDmicroscopy2015,CSDSuperresMagImg2019} produces $\rm{NV^-}$ to $\rm{NV^0}$ photoionization under the doughnut, which makes them less relevant to sensing; and spin-RESOLFT \cite{SpinRESOLFT_Early_2010,SpinRESOLFT2017} eliminates the sensing information stored in all NV spins except those in the doughnut  center after the sensing protocol has been applied (and before spin-state readout). A second class of methods uses stochastic superresolution microscopy techniques such as STORM \cite{STORM_Englund2013,STORM_Wrachtrup2014} and PALM. However, stochastic methods appear challenging to use in high-NV density regimes. Finally, gradients of external magnetic fields can be used to achieve sub-diffraction-limit resolution \cite{FourierImaging2015_Arai,FourierSelectiveEnsemble2017,FourierSelectiveWalsworth2017,Bgradient_HardDriveWrachtrup2018}. A large magnetic gradient over a small region in diamond induces a strong spatial variation of Zeeman splittings in NVs distributed within that region; this allows one to spatially resolve NVs using a tunable MW frequency source. Magnetic gradients are also used in Fourier magnetic imaging, which uses pulsed magnetic field gradients to encode the position information of an ensemble of NVs in a spatial modulation of their spin coherence phase (i.e., in spatial Fourier or k-space); this spatial information is not limited by optical diffraction and a real-space image can be extracted via Fourier transform after acquiring the data. Similarly, spatial MW gradients have also been shown to enable sub-diffraction-limit resolution of NVs \cite{inhomogeneous_MW_eng_2019}. In Table\,\ref{tbl:superresolution}, we summarize the results reported on superresolution techniques in NV-diamond systems.

\begin{table*}[htbp]
\centering
\begin{tabular}{l|c|c|c}
Method  & Resolution &  Modality & Ref.\\
\hline\hline 
STED & $\sim 10$ nm & {Real space, Scanning} & \cite{STED_Early_Hell2009}\\
SIL-STED & $\sim 2.5$ nm & {Real space, Scanning} & \cite{STED_SIL2012}\\
CSD & $\sim 4$ nm & {Real space, Scanning} & \cite{CSDmicroscopy2015}\\
spin-RESOLFT & $\sim 20$ nm & {Real space, Scanning} & \cite{SpinRESOLFT2017}\\
SAM & $\sim 17$ nm & Real space, Scanning & \cite{AiryDisc_Gardill:2022mxg}\\
STORM & $\sim 25$ nm & {Real space, Widefield} & \cite{STORM_Wrachtrup2014}\\
Magnetic gradient & $\sim 30$ nm &  {Fourier space, Widefield} & \cite{FourierImaging2015_Arai}\\
Magnetic gradient & $\sim 30$ nm &  {Real space, Widefield} & \cite{FourierSelectiveWalsworth2017}\\
MW gradient & $\sim 100$ nm &  {Real space, Widefield} & \cite{inhomogeneous_MW_eng_2019}
\end{tabular}
\caption{\label{tbl:superresolution} Superresolution techniques demonstrated in NV-diamond sensing. STED: stimulated emission depletion. SIL: solid immersion lenses fabricated directly in the diamond. CSD: charge state depletion. spin-RESOLFT: spin measurement using reversible saturable optical linear fluorescence transitions. STORM: stochastic optical reconstruction microscopy. SAM: Super-resolution Airy disk Microscopy.}
\end{table*}

All of the methods discussed achieve or approach the benchmark resolution of 20 nm needed to detect directionally oriented WIMP signals. The demonstrations, however, have been limited to thin NV layers or samples with sparse NV centers, passively imposing 3D resolution. In order to adapt these techniques for WIMP track reconstruction, it is important to integrate depth-resolution into any of the methods. The 3D analogue of doughnut-shape illumination has been attempted, reaching about 100 nm z-resolution \cite{STEDHell2009_3D,STED2016_Mirror}. However, the high index of refraction of diamond might impede further progress in this route for achieving 3D resolution of 20 nm.

\begin{figure}[htbp]
\begin{center}
\includegraphics[width=0.48\textwidth]{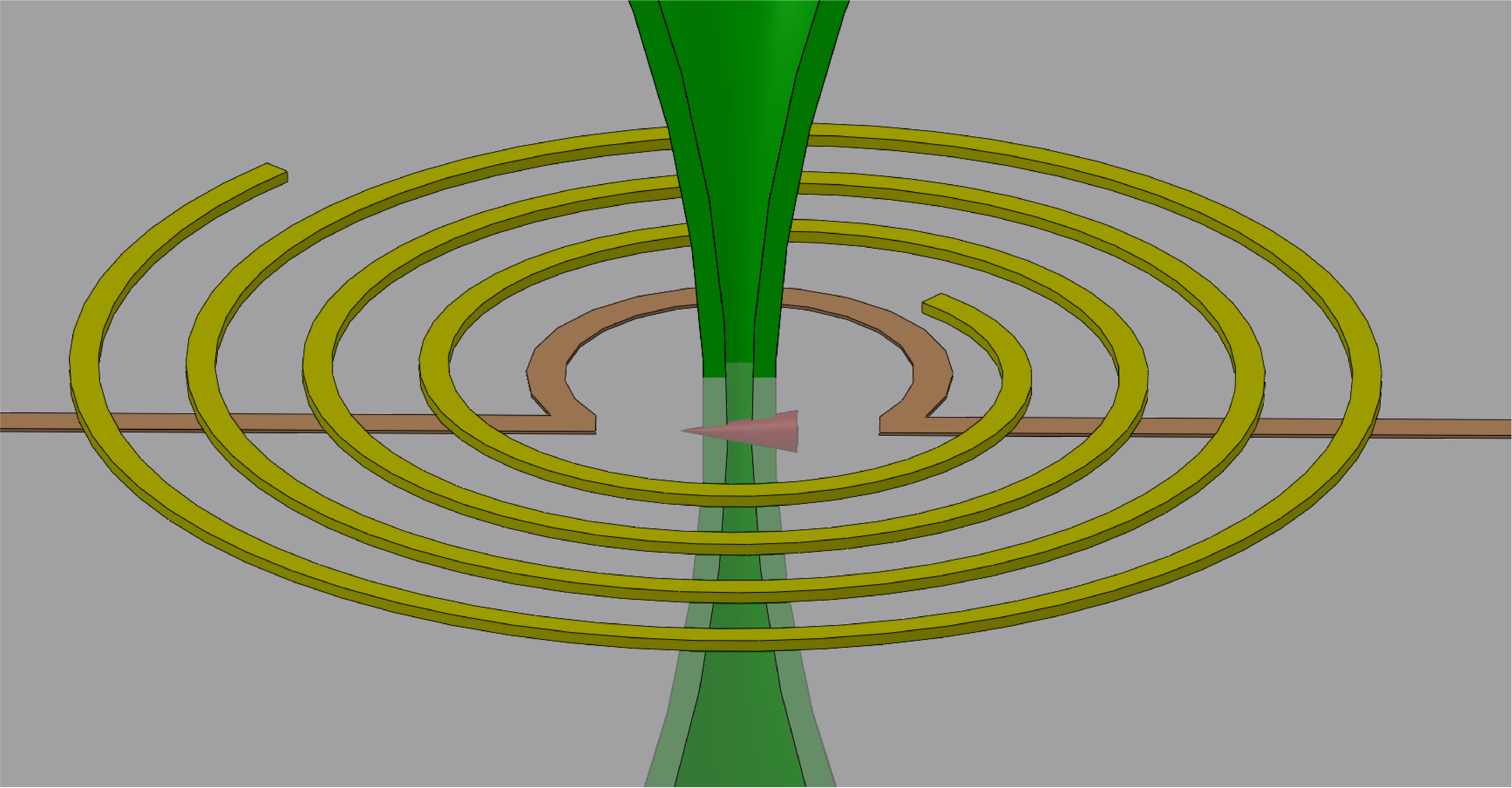}
\end{center}
\caption{\label{fig:superresolution} Illustration of the combined method proposed for superresolution NV strain spectroscopy. A doughnut-shaped illumination provides lateral resolution through STED, CSD, or spin-RESOLFT. Fabricated gradient coils produce magnetic field gradients that provide depth resolution. An $\Omega$-shape waveguide for microwave delivery is also illustrated. From M. C. Marshall, M. J. Turner, M. J. H. Ku, D. F. Phillips, and R. L. Walsworth, Quantum Sci. Technol. 6, 024011 (2021). Reproduced by permission of IOP Publishing Ltd.}
\end{figure}

For the reconstruction of neutrino or WIMP-induced strain (damage track) in three dimensions, we propose a combined technique: STED, CSD, and spin-RESOLFT methods can be used for achieving lateral resolution, while simultaneous magnetic field gradients can be used for achieving depth resolution \cite{Marshall:2020azl}. Figure\,\ref{fig:superresolution} illustrates this scheme. Using this combined method, it is estimated that it will take about 11 hours to scan a 1 $\rm{\mu m}^3$ volume with 10 nm resolution in all three dimensions \cite{Marshall:2020azl}. In addition, to avoid complications of mapping out the damage track deep in the diamond, fast and low-strain etching methods \cite{strain_control_2009_friel,Etching_fast_2018_Nagai,Etching_highQ_fast_2019_Hicks} can be employed to have the damage track closer to the surface instead of deep in the bulk. Spectroscopic methods such as spin-RESOLFT and Fourier imaging require good quantum coherence properties that are present in CVD diamond but not commonly in HPHT diamond samples. Therefore, the proposed combined method may not be applicable to HPHT samples with high nitrogen content in the detection scheme based on NV creation (see Sec.\,\ref{sec_FNTD}). However, due to low initial density of NVs, it will be possible to directly resolve created NV centers. In summary, state of the art in superresolution imaging using NV centers suggests a promising path for 3D damage reconstruction in either scheme of the detector, i.e., localization using strain spectroscopy (see Sec.\,\ref{sec_strain_spectroscopy}) or localization using creation of NV centers (see Sec.\,\ref{sec_FNTD}).

\section{Sensitivity characterization via single ion implantation}
\label{sec_ion_implantation}

Thus far, we have discussed methods that enable the localization of WIMP and neutrino damage tracks and retrieval of the directional information encoded in the track geometry and orientation. Here, we present a method to generate injected signals that can be used to determine the directional sensitivity of the detector.

State-of-the-art focused ion beams (FIB) offer single ion number resolution, low-energy ions, high spatial resolution, as well as high mass resolution \cite{ioncounting_2017_pacheco_implanter_details}. In the ion implanter at Ion Beam Laboratory (IBL) at Sandia National Laboratories (SNL), the ions are extracted from liquid metal alloy ions sources (LMAIS), which allows working with a third of the atoms in the periodic table. Electrostatic ExB filters incorporated in the setup provide mass resolution of $M/\Delta M>61$. A fast beam-blanker is implemented with a minimum pulse time of about 16\,ns. Lastly, the electrostatic objective lens provides submicron ion beam spots \cite{ioncounting_2017_pacheco_implanter_details}. The combination of these features makes this setup a suitable platform for generating single-particle-induced damage tracks in diamond, simulating WIMP or neutrino-induced recoil tracks. 

\begin{figure}[htbp]
\begin{center}
\includegraphics[width=0.48\textwidth]{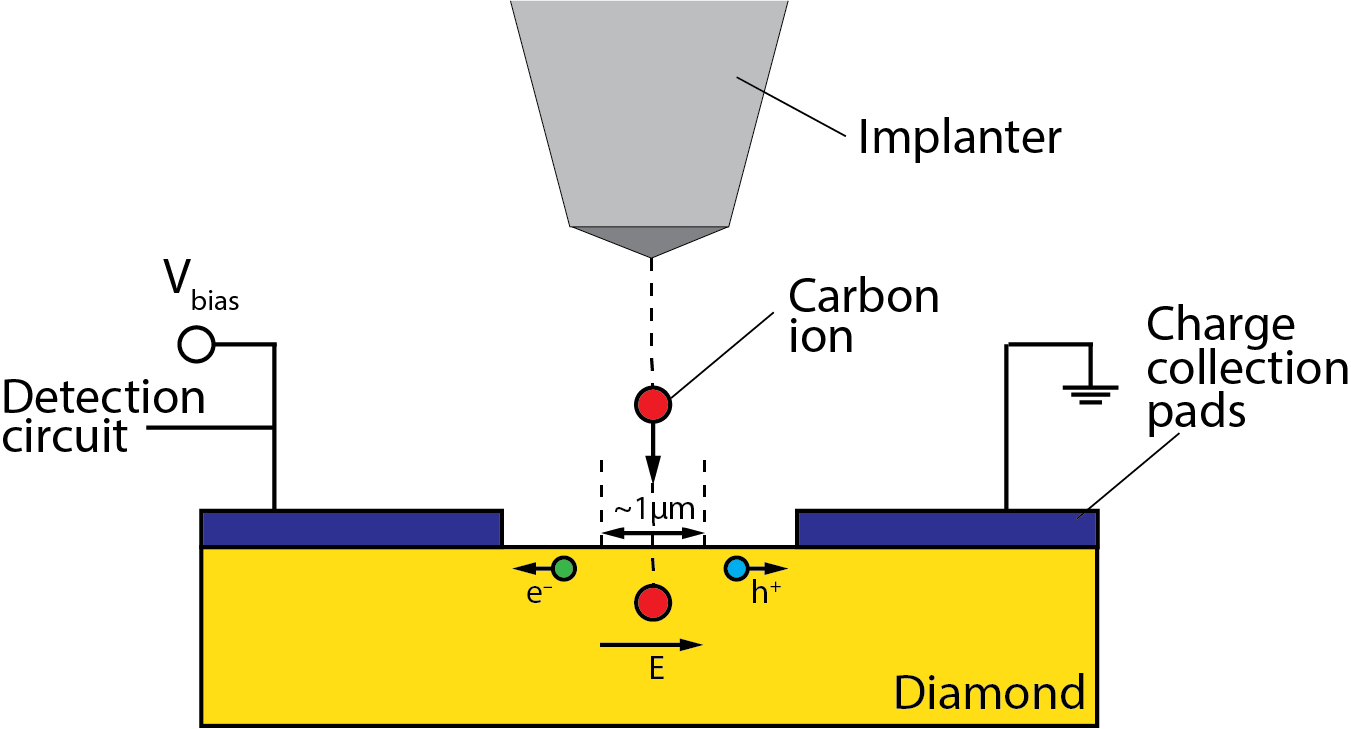}
\end{center}
\caption{\label{fig:ion_implantation} Illustration of the single ion implantation experiment using in situ ion counting method. The produced electron-hole pairs are collected under a DC bias voltage through the fabricated pads.}
\end{figure}

With fast beam blanking, pulsed ion implantation can achieve low ion numbers. In spite of this, as the number of ions in a fixed pulsed time is determined by Poisson statistics, there is large relative uncertainty regarding the number of ions in a pulse at low ion counts. In situ ion counting is a method developed to overcome the Poissonian uncertainty \cite{ioncounting_2005_jamieson_controlled,ioncounting_2008_seamons,ioncounting_2010_Bielejec,ioncounting_2016_abraham_fabrication,ioncounting_2017_pacheco_implanter_details,Titze:2021xpf}. In this method, illustrated in Figure\,\ref{fig:ion_implantation}, metallic pads with a separation of about 10 microns are fabricated on the sample surface. Ion beams are focused on the gap between the pads. During the implantation, the pads are biased with a DC voltage and are connected to a collection circuit for monitoring the charge generation due to ion implantation, i.e., ion-induced electron-hole production. The following method uses a charge collection system to achieve precise single ion implantation. Based on the beam-off charge readout, a lower threshold for the collection circuit voltage is defined. The mean ion number per pulse during the implantation is set to a small number (e.g., $\braket{\rm ion~number/pulse}\lesssim0.1$). Poisson statistics suggests that there are mostly zero ions per pulse, but occasionally there are one or more (but mostly one ion). As the implantation continues pulse by pulse at the same site, the readout voltage is monitored and compared to the pre-determined threshold. As soon as the voltage passes the threshold, implantation is stopped at that site. Besides threshold-based implantation, the full distribution of ion-induced voltage can also be used in post-processing to reduce the ion-number uncertainty \cite{Titze:2021xpf}. A recent example is the use of this method for silicon implantation on diamond with a 5\% ion number error \cite{Titze:2021xpf}.

We propose using single carbon ion implantation experiments to generate WIMP-like signals in diamond. The implanted ion will interact with native diamond lattice carbon atoms initiating nuclear recoils, much like the initial recoil caused by WIMP or neutrino interactions. The single-ion implantation can be used to test both methods of localization described in Sections \ref{sec_strain_spectroscopy} and \ref{sec_FNTD}. For NV-based strain spectroscopy, the implantation will take place in a high-quality CVD diamond with an ensemble of NV centers. The ion-induced strain signal can then be examined and detector efficiency can be determined as a function of implantation energy (corresponding to recoil energy, thus the WIMP mass). For NV creation, an HPHT diamond sample with low NV density and high nitrogen density will be used. During high-temperature annealing, some ion-induced vacancies will combine with the nitrogen atoms, forming NV centers that can be detected with NV fluorescence microscopy. The use of such an experimental method will provide insight into NV creation efficiency and the effects of annealing on initial damage track distortion.

%% file: conclusion.tex
\section{Summary and outlook}
\label{sec_discussion}

Results from upcoming generations of WIMP detectors are likely to approach the ``neutrino floor'', where coherent scattering of low-energy neutrinos will be detected \cite{darwin_2016,argo_2018,solar_nu_search_xe_2021}. Like WIMPs, neutrinos induce nuclear recoils in a target. They scatter at energies relevant to WIMP searches, meaning standard background discrimination techniques cannot reject them. Without discrimination between WIMP and the neutrino events, identifying WIMPs below the neutrino floor will require detecting annual modulation atop the neutrino background, demanding several dozen events to achieve a five-sigma discovery of a new particle. A directional particle WIMP detector could reject the solar neutrino background in the early stages of operation \cite{directional_2014,directional_2015,Mayet:2016zxu,Vahsen:2020pzb,Vahsen:2021gnb}, and then other astrophysical and terrestrial neutrinos in the later stages as they become relevant at higher exposures. Further, a directional detector could reveal a WIMP signal’s cosmological origin, and potentially improve our understanding of local dynamical DM structures \cite{directdetection_MW_debrisflow_2012,OHare:2018trr}. Solid-state directional detectors are especially attractive for probing WIMP cross-sections below the neutrino floor. Their high target density contrasts well to existing directional detection methods with gaseous and emulsion targets. In addition, the proposed solid-state detector using quantum defects in diamond would provide timing information as well as complete three-dimensional information via head/tail asymmetric signals \cite{Rajendran:2017ynw,Marshall:2020azl} (similar to the well-developed gas TPC detectors.)

\textbf{Detector principle.}
Wide-bandgap semiconductors such as diamond \cite{Kurinsky:2019pgb} and silicon carbide \cite{Griffin:2020lgd} have been proposed as a target for solid-state WIMP detection. Their good semiconductor properties and lower-mass nuclei provide an advantageous sensitivity profile compared to existing detectors. These materials can be lab-grown with high purity and homogeneous crystal structure \cite{strain_control_2009_friel,strain_control_2009_friel,strain_lowdislocations_typeIIa_2009_martineau_high}; a WIMP event in such a crystal would leave a characteristic track of damage, with the crystal acting as a “frozen bubble chamber” recording the direction of the incident particle \cite{Rajendran:2017ynw,Marshall:2020azl}. The crystal damage track results from the cascade of secondary nuclear recoils initiated when a WIMP (or neutrino) impacts a target nucleus; simulations for a diamond target indicate measurable orientation and head/tail asymmetry down to 1–3 keV of recoil energy \cite{Rajendran:2017ynw}. The shape and orientation of this damage track can be read out via spectroscopy of quantum point defects in the crystal such as nitrogen-vacancy (NV) \cite{NV_review_2014_Schirhagl,NV_review_Levine_2019,NV_review_2020_edmonds,sensitivity_Barry2020} and silicon-vacancy (SiV) \cite{SiV_2018} color centers in diamond, and divacancies \cite{divacancy_SiC_2018_Awschalom,divacancy_SiC_2018_Gali} in silicon carbide.

A directional detector based on solid-state point defects in a semiconductor could take advantage not only of a large target mass, but also of intensive development of instrumentation for WIMP detectors based on silicon or germanium \cite{SuperCDMS_LopezAsamar:2019smu,SuperCDMS_Rau:2020abt,DAMIC:2021crr,DAMIC:2021esz}. Detector segments could be instrumented with charge or phonon collectors such as transition edge sensors, or scintillation photons could be collected from the detector bulk. When a crystal segment triggers one of these detection methods, the event time would be recorded and the segment would be removed from the detector for directional analysis, while the remainder of the detector continues to accumulate exposure (Sec.\,\ref{sec_detector_principle}).

Damage tracks from WIMP and neutrino events will be tens or hundreds of nanometers long \cite{Rajendran:2017ynw}, requiring a two-step process to extract the directional information \cite{Marshall:2020azl}. Fortunately, both steps can be built upon techniques established in the fields of solid-state quantum sensing and quantum information processing. First, diffraction-limited optics can be used to resolve the position of the damage track to a sub-micron voxel within the millimeter-scale detector segment (sections \ref{sec_strain_spectroscopy} and \ref{sec_FNTD}). Second, optical superresolution techniques (Sec.\,\ref{sec_superresolution}) and/or high-resolution x-ray nanoscopy (Sec.\,\ref{sec_xray_microscopy}) can be used to measure the damage track structure at the nanometer scale.

\textbf{Directional readout technologies.} Initial simulations and experimental demonstrations of these techniques have focused on NV centers in diamond, as the NV is the most widely used and best-characterized quantum defect \cite{sensitivity_Barry2020}. Consisting of a lattice site vacancy adjacent to a substitutional nitrogen defect, the NV is an electronic spin-1 system featuring optical initialization and readout of the spin state, and microwave transition frequencies sensitive to local crystal strain (as well as magnetic and electric fields, and temperature); see Sec.\,\ref{sec_nv}. The damage track from a neutrino or WIMP-induced recoil cascade would induce significant strain on nearby NV centers, which can be measured via shifts in their spin transition frequencies. Diffraction-limited strain imaging (Sec.\,\ref{sec_strain_spectroscopy}) could be used to localize damage tracks at the micron scale; sensitive, widefield strain imaging has been the subject of much recent work \cite{Kehayias_2019,Broadway_2019,marshall_2021_strainCPMG}. Superresolution microscopy and spectroscopy using NV centers (Sec.\,\ref{sec_superresolution}) has been extensively developed as well. Spatially-resolved, subdiffraction strain sensing is a plausible avenue for nanoscale readout of damage track direction. These same methods are applicable to other color centers in diamond such as the silicon-vacancy center, or to divacancies in silicon carbide.

Alternatively, quantum defects can be created from recoil-induced lattice site vacancies generated during a WIMP or neutrino interaction with the diamond (Sec.\,\ref{sec_FNTD}). For example, a nitrogen-rich diamond could be annealed to induce these vacancies to form new NV centers or other color centers, highlighting the damage track. In crystals with few pre-existing emitters, this represents a near-background-free method of damage track localization. The direction can then be extracted either via strain spectroscopy of the created color centers, or by superresolution spatial mapping of the defect positions. Scanning X-ray nanobeam diffraction measurements (Sec.\,\ref{sec_xray_microscopy}) present a complementary, non-defect-based method for damage track direction measurement – instruments at synchrotron facilities can detect strains at the level predicted for a WIMP damage track, and can be performed with nanometer spatial resolution.

All of the techniques outlined above require that optical measurements be performed on-site to avoid cosmic ray exposure during transit. The low cost and simplicity of the setups for these measurements enable multiple instruments to be used in parallel. Additionally, detector segments may require etching to bring the damage track close to the surface for nanometer-scale measurements, but appropriate techniques should enable this process without introducing additional strain or distorting the WIMP signal \cite{Etching_highQ_fast_2019_Hicks,strain_control_2009_friel}.

\textbf{Outlook.} In the near term, the work towards such a solid-state WIMP detector with directional sensitivity will be centered around demonstrating the capability to locate and determine the direction of nuclear recoil damage tracks in diamond. This effort requires adaptation and development of existing techniques, but the current state of the art is not far from the requisite sensitivity and resolution. In particular, our recent work on widefield strain imaging in diamond via NV spectroscopy \cite{marshall_2021_strainCPMG}, as well as scanning X-ray diffraction microscopy of nanoscale strain mapping \cite{marshall_2021_xray}, demonstrate the sensitivity required for directional detection. For concrete experimental characterization of the efficiency of these methods, injected nuclear recoil signals can be generated using single ion implantation (Sec.\,\ref{sec_ion_implantation}).

In the medium term, instrumentation of a prototype detector will be required. Such a prototype would feature low-threshold charge, phonon, or photon collection capabilities with spatial resolution at the millimeter scale, as well as the micron and nano-scale strain mapping capabilities discussed above. Development of crystal growth techniques to create large volumes of structurally homogeneous crystals is also required. Modern diamond growth techniques using chemical vapor deposition (CVD) enable repeatable, fast, and low-cost growth of uniform crystals; reasonable progress in this field, expected in the next decade, should allow for the production of the volume of diamond material required for WIMP detection. With appropriate development and synergy between multiple disciplines, the proposed approach offers a viable path towards directional WIMP detection with sensitivity below the neutrino floor.